\documentclass[preprint2]{aastex}

\usepackage{epsfig}
\newcommand\etal{{\it et al}}                                                  
\def\deg{^{\circ}}

\received{2001 November 9}
\begin{document}
\title{Toward An Empirical Theory of Pulsar Emission. VII. On the 
        Spectral Behavior of Conal Beam Radii and Emission Heights}
\shorttitle{An Empirical Theory of Pulsar Emission. VII}
\shortauthors{Mitra \& Rankin}

\author{Dipanjan Mitra} 
\affil{Max-Planck Institut f\"ur Radioastronomie, Bonn, Germany; 
email: dmitra@mpifr-bonn.mpg.de}
\and
\author{Joanna M. Rankin\footnote{On leave from the Physics Department, 
University of Vermont, Burlington, VT 05405 USA; email: 
rankin@physics.uvm.edu}}
\affil{Sterrenkundig Instituut `Anton Pannekoek', Universiteit van Amsterdam, 
Amsterdam 1098 EL NETHERLANDS; email: jrankin@astro.uva.nl}

\begin{abstract}
In this paper we return to the old problem of conal component-pair 
widths and profile dimensions.  Observationally, we consider a set 
of 10 pulsars with prominent conal component pairs, for which well 
measured profiles exist over the largest frequency range now possible.  
Apart from some tendency to narrow at high frequency, the conal 
components exhibit almost constant widths.  We use all three profile 
measures, the component separation as well as the outside half-power 
and 10\% widths, to determine conal beam radii, which are the focus 
of our subsequent analysis.  These radii at different frequencies 
are well fitted by a relationship introduced by Thorsett (1991), 
but the resulting 
parameters are highly correlated.  Three different types of behavior 
are found:  one group of stars exhibits a continuous variation of 
beam radius which can be extrapolated down to the stellar surface 
along the ``last open field lines''; a second group exhibits beam 
radii which asymptotically approach a minimum high frequency value 
that is 3--5 times larger; and a third set shows almost no spectral 
change in beam radius at all.  The first two behaviors are associated 
with outer-cone component pairs; whereas the constant separation 
appears to reflect inner-cone emission.  The first group, remarkably, 
can be fitted by a Thorsett relation in which the constant term is 
constrained to be the field-tangent direction at the edge of the 
polar cap $\rho_{\rm pc}$, but the others cannot.  The first group 
can also be fitted well using an index of --1/3, but the second 
group cannot.

We first compute heights from the conal beam radii, assuming dipolar 
fields and emission along the ``last open field lines''---which we 
find are again well fitted by a suitable Thorsett relation.  Here we 
find that the first group can be fitted using a constant term 
$h_{\rm pc}$ of 10 km and also that the first two groups are 
remarkably well fitted by an index of --2/3.  We then argue that 
physical emission heights can be estimated using the component 
separation along an interior annulus of field lines having their 
``feet'' about halfway out on the polar cap---and such values agree 
well with most existing height values based on physical criteria.  

Therefore, we find that ``radius-to-frequency'' mapping is associated 
with outer-conal component pairs.  The near constant behavior of 
inner-cones is thus arresting.  We explore possible interrelationships 
between the spectral behavior of the component and profile widths 
produced both by the field-line flaring and the changing sightline 
geometry.  We also attempt to understand the physical implications 
of the parameter values resulting from the Thorsett-relation fits.  
\end{abstract}


\section{Introduction}

Those few bright, relatively local pulsars exhibiting 
a well resolved pair of conal components have had major 
influence on our conception of pulsar radio emission.  
They represent only a handful of the 1000+ stars now 
known, and some of their original ``B'' designations 
come easily to the tongue: B0301+19, B0329+54, B0525+21, 
B1133+16, B1237+25, B2020+28, and B2045--16, having each 
been the subject of sundry studies over the years.  
This small group comprises the finest examples of stars 
exhibiting a monotonic increase of total profile width, 
or conal component-pair spacing, with wavelength.  Such 
behavior, noted in the earliest efforts to systematize 
and understand pulsar radio emission ({\it e.g.} 
Komesaroff 1970, Komesaroff, Morris \& Cooke 1970, Lyne, 
Smith \& Graham 1970; Sieber, Reinecke \& Wielebinski 
1975; Cordes 1978), has generally been regarded as 
primary evidence that the lower frequency radio 
emission is emitted at higher altitudes in the pulsar 
magnetosphere.  While this phenomenon, known as 
``radius-to-frequency mapping'' (hereafter RFM), is 
clearly exhibited by many pulsars, it is particularly 
marked in lower frequency ($f$$<$100 MHz) observations.  
Consequently, it is only a few local (low dispersion) 
objects that can be observed over the 8--10-octave 
frequency range needed to examine its character in 
adequate detail.  

Thorsett (1991) nicely summarizes the early history of 
the RFM discussion:  Observationally, it rapidly became  
clear that no simple curve fitted the measurements, so 
that the component-pair width (or separation) behavior 
(as longitude intervals $\Delta\varphi$, 360$\deg$ 
corresponding to one rotation period $P$) was usually 
described by fitting power-law functions $f^{-a}$ to the 
asymptotic high and low frequency values.  Theorists, 
meanwhile, began to predict various dependences for the 
RFM---the height proportional to $f^{-2/3}$ relation 
encountered in models where the characteristic emission 
frequency is equated to the plasma frequency (such as 
for curvature radiation)---being the most often 
encountered.  Later studies suggested that the index 
$a$ varied continuously with frequency ({\it i.e.} 
Rankin 1983b; hereafter Paper II) and, therefore, that 
no single power-law relation could describe the overall 
behavior (Slee, Bobra \& Alurkar 1987).  Thorsett then 
evaluated two simple generalizations of a single 
power-law
\begin{equation}
\Delta\varphi = (f/f_o)^{a} + \Delta\varphi_{\rm min}{\rm ,\ \ and} 
\label{eq1}
\end{equation}
\begin{equation}
(\Delta\varphi)^2 = (f/f_o)^{a} + (\Delta\varphi_{\rm min})^2, 
\end{equation}
(where $f$ is the radio frequency) and found that they 
fitted his measurements equally well.  

Over this same period, ideas developed regarding the 
geometrical significance of these prominent component 
pairs.  First, they were demonstrably {\it conal}---that is, 
they represented a relatively central sightline traverse 
through a hollow-conical emission pattern (Rankin 1983a; 
hereafter Paper I).  Second, the magnetic latitude $\alpha$ 
and sightline impact angle $\beta$ could be estimated by 
various techniques ({\it e.g.} Lyne \& Manchester 1988), 
taking the polarization-angle (hereafter PA) sweep rate 
$R$ to be $|\sin\alpha/\sin\beta|$ (Radhakrishnan \& Cooke 
1969; Komesaroff 1970).  And third, these conal beam radii 
assumed particular dimensions in terms of the polar cap 
radius, which in turn suggested specific emission heights 
(Rankin 1993a,b; hereafter Paper VIa,b).  

Surprisingly, few attempts have been made to amalgamate 
these discussions, to the end that questions relating to 
the frequency dependence of width (or separation) can be 
explored in the context of a specific emission geometry.  
Two such attempts are those of Xilouris \etal\ (1996) 
and von Hoensbroech \& Xilouris (1997b) which we will 
discuss further below.  Clearly, a much more appropriate 
quantity with which to assess the RFM is the conal beam 
radius $\rho$ (which has primary geometrical significance; 
see Paper VIa: fig. 2), rather than the various observed 
intervals of longitude $\Delta\varphi$ (which conflate 
the accidental geometry of the observer's sightline).  
Moreover, once $\rho$=$\rho(f)$ has been determined, the 
emission height $h$ can easily be estimated (assuming a 
dipolar magnetic field geometry), so that the RFM can be 
assessed directly as a frequency-dependent height relation 
$h$=$h(f)$.  Our purpose in the remainder of this paper 
is then a) to combine measurements used by Thorsett with more 
recent and accurate profile observations, b) to study 
their implications in the context of an appropriate 
geometrical model, and c) to draw what physical 
conclusions we can from the results of these analyses. 

\notetoeditor{Table 1 should be a single-column table.  If it does not fit 
as a single column, we would be willing, but not eager, to have the second 
column, the pulsar period, omitted.}

\section{Measurements and Analysis}

\begin{deluxetable}{cccccccc} \tablewidth{0pt} \tablecaption{Pulsar
Parameters.\label{tbl-1}} \tablehead{Pulsar &Period & DM & $f_{\rm 1\deg}$
&B$_{12}$ &$\alpha$ & $R$ &  $\beta/\rho$ \\
  (B--) & (s) & (pc~cm$^-3$)& (MHz) & (G) & ($\deg$)& & [at 1GHz] } 
\startdata
 0301+19 & 1.388 & 15.69 & 41 & 2.7  &$38\pm5$ &$17\pm1$ & 0.45 \\
 0525+21 & 3.745 & 50.87 & 42 & 24.6 &$21\pm2$ &$36\pm1$ & 0.19 \\
 1237+25 & 1.382 & 9.28  & 47 & 2.34 &$53\pm3$ &$\infty$?&$\sim$0\\
 2045--16& 1.962 & 11.51 & 40 & 9.34 &$34\pm2$ &$30\pm2$ & 0.26 \\[4pt]
 0329+54 & 0.715 & 26.78 & 103& 2.46 &$32\pm3$ &$10\pm3$ & 0.31 \\
 1133+16 & 1.188 & 4.85  & 41 & 4.26 &$46\pm3$ &$10\pm1$ & 0.78 \\
 2020+28 & 0.343 & 24.6  & 164& 1.62 &$56\pm5$ &$6\pm1$  & 0.49 \\[4pt]
 0834+06 & 1.274 & 12.86 & 55 & 5.9  &$30\pm6$ &$9\pm2$  & 0.86 \\
 1604--00& 0.422 & 10.72 & 108& 0.72 &$50\pm5$ &$8\pm2$  & 0.82 \\
 1919+21 & 1.337 & 12.43 & 53 & 2.7  &$34\pm3$ &$11\pm2$ & 0.76 \\ \enddata
\end{deluxetable}  

Table~\ref{tbl-1} gives parameters of the pulsars which 
are the subjects of this analysis---that is, two groups 
comprising the seven primary stars already mentioned as 
well as three others---which will be discussed as needed 
below.  Our profile measurements were taken both from 
those sources identified by Thorsett (see his Table I and 
the references therein)  and from more recent published 
and unpublished  work.  These additional sources are 
listed below Table~\ref{tbl-2} and include high quality 
Arecibo observations at between 25 MHz and 6 cm (Hankins, 
Rankin \& Eilek 2002).  

The various observations were reduced to sets of profile 
measurements using a combination of hand-scaling and, 
where appropriate, fitting Gaussian functions to the 
relevant components (see Kramer 1994; Kramer \etal\ 1994).  
In all cases we were interested in a) the half-power (3-db) 
widths of the respective components, b) the separations 
between the centers of the component pairs, c) the full 
profile widths at the outside half- and 1/10-power points, 
and d) reasonable error estimates of these values.  In 
practice the outside half-power (3-db) points were determined 
separately for the two components as their amplitudes were 
usually different (see Backer 1976; figs. 2c,d).  Generally,  
we found the fitting more useful to locate the component 
centers for the separation measurements, but in many cases 
the 3-db profile widths could also be accurately estimated 
by summing the component-pair separation and half the sum 
of their fitted widths.  As can be seen in the following 
plots and tables, it was usually possible to estimate a 
given quantity in several different ways (including making 
use of Thorsett's independent values), so that there is a 
good deal of redundancy in our measure values. 

The errors in the widths $\sigma_{\Delta\varphi}$ were 
estimated wherever possible by using the formal errors 
of the Gaussian fits, which it turn were based on well 
determined values of the noise in the off-pulse region 
of the profile.  Where hand-scaling was used, the errors 
were again estimated from the quality of the profile and 
the off-pulse rms-noise level.

\section{Component-Width Spectra}

\begin{figure}
\begin{center}
\epsfig{file=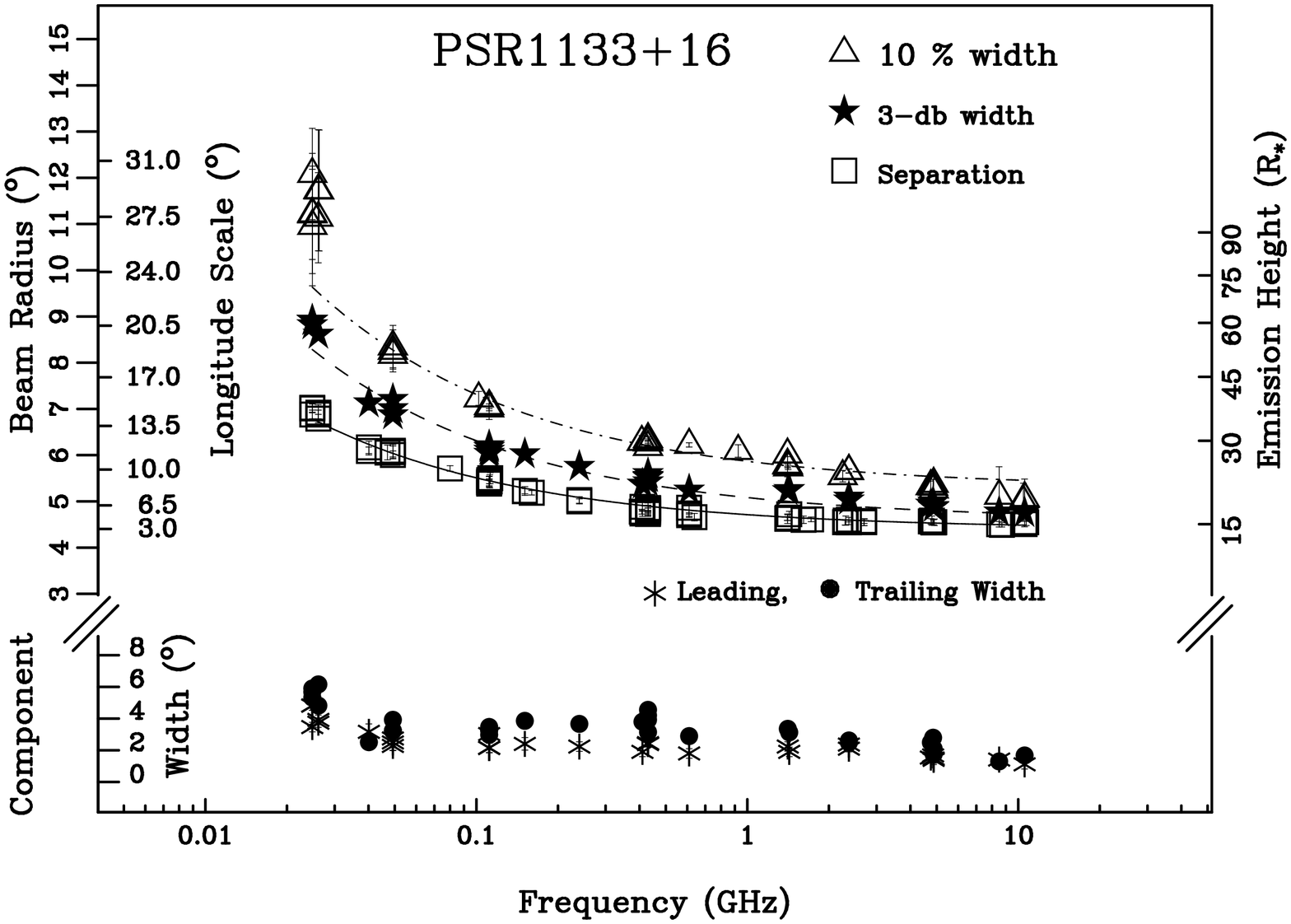,height=8cm,width=6cm,angle=-90}
\end{center}\caption[]{Component-pair widths (lower portion), 10\% 
and 3-db profile widths, and component separations (upper 
portion) for pulsar B1133+16 as a function of frequency.  
The component and profile widths have been corrected for 
the estimated low frequency instrumental broadening (see 
text), significant here only at 25 MHz where it has 
reduced the plotted values by around 1$\deg$. The three 
profile measures are plotted according to three separate 
scales: longitude (inside left), inferred beam radius 
(outside left), and inferred emission height (outside 
right); note that the first and last scales are 
necessarily somewhat non-linear.  Keys to the respective 
symbols are given on the plot.  The respective curves, 
fitted both to the conal beam radii and component widths, 
are discussed in the text.} 
\label{fig:1133wdt}
\end{figure}

Most previous attempts to understand how the widths 
of profile components vary with frequency have only 
proved frustrating.  Until the last decade or so, no 
pulsar's profiles were well enough measured---particularly 
at low frequency---that this important issue could be 
addressed with any confidence.   As the brightest star 
in the sky with two well resolved profile components, 
pulsar B1133+16 ostensibly offers the best possible 
context for studying its component-width behavior, but 
Rankin's (Paper II) effort to draw conclusions from the 
then available profile data (see her fig. 5), we will now 
see, was inconclusive and misleading in its suggestion 
of a monotonic increase of width with wavelength---though 
Izvekova \etal\ (1993), a decade later, were able to 
discern a nearly invariant behavior over a narrower 
frequency range.  

The lower portion of Figure~\ref{fig:1133wdt} gives a 
pair of component-width spectra for B1133+16, where the 
values for the leading (I: asterisks) and trailing (II: 
circles) components are plotted according to the 
``Component Width'' scale.   While these width values 
are hardly constant, they do not begin to mimic the 
monotonically varying profile-width curves in the upper 
part of the figure.  Though we see discrepancies among 
the well measured values ({\it e.g.,} around 400 MHz), most 
of the values are roughly compatible, within their errors, 
with a constant width at frequencies above about 40 MHz: 
2.2$\deg$$\pm$0.6$\deg$ and 3.2$\deg$$\pm$0.9$\deg$ for 
components I and II, respectively.  Only below 40 MHz 
and around or above 3 GHz do the component-widths 
diverge from these values.

Further, the increases below 50 MHz appear to be just 
what might be expected from instrumental effects.  
Scattering broadening also becomes important for these 
low dispersion measure $DM$ objects at low frequencies, 
but it is usually quite clear when it is the dominant 
effect, because of its steep $f^{\sim-4.4}$  dependence.  
At the lowest frequencies  the necessarily narrow filter 
bandwidths $\Delta f$ entail risetimes which become 
comparable to the dispersion smearing---so that the 
minimum achievable (non-coherently dedispersed) 
resolution escalates as $f^{-3/2}$.  If the filter 
risetime $\Delta t_r$ is of the order of $2\pi k/\Delta 
f$ (where $k$ is a small positive value), and the 
dispersion smearing $\Delta t_d$ goes as ${DM}\Delta f/f^3$, 
then for optimum choices of parameters $\Delta t_{\rm min}$ 
($=\Delta t_r = \Delta t_d$), we find that $\Delta t_{\rm min} (\deg)$ 
is about 145$\deg[k{DM}/P^2(\rm s)]^{1/2}f(\rm MHz)^{-3/2}$.   
Table~\ref{tbl-1} gives the frequencies $f_{\rm 1\deg}$ 
at which this limiting resolution represents a longitude 
of $1\deg$ (for $k$=1).  Were then $k$ very reasonably 
some 2--4, we can understand the width escalation for 
1133+16 at and below 50 MHz as instrumental in origin.  
We have thus corrected the low frequency width values 
for this limited resolution (taking $k$ to be about 3) 
on the plots and in our subsequent analyses.  

\begin{figure*}
\begin{center}
\epsfig{file=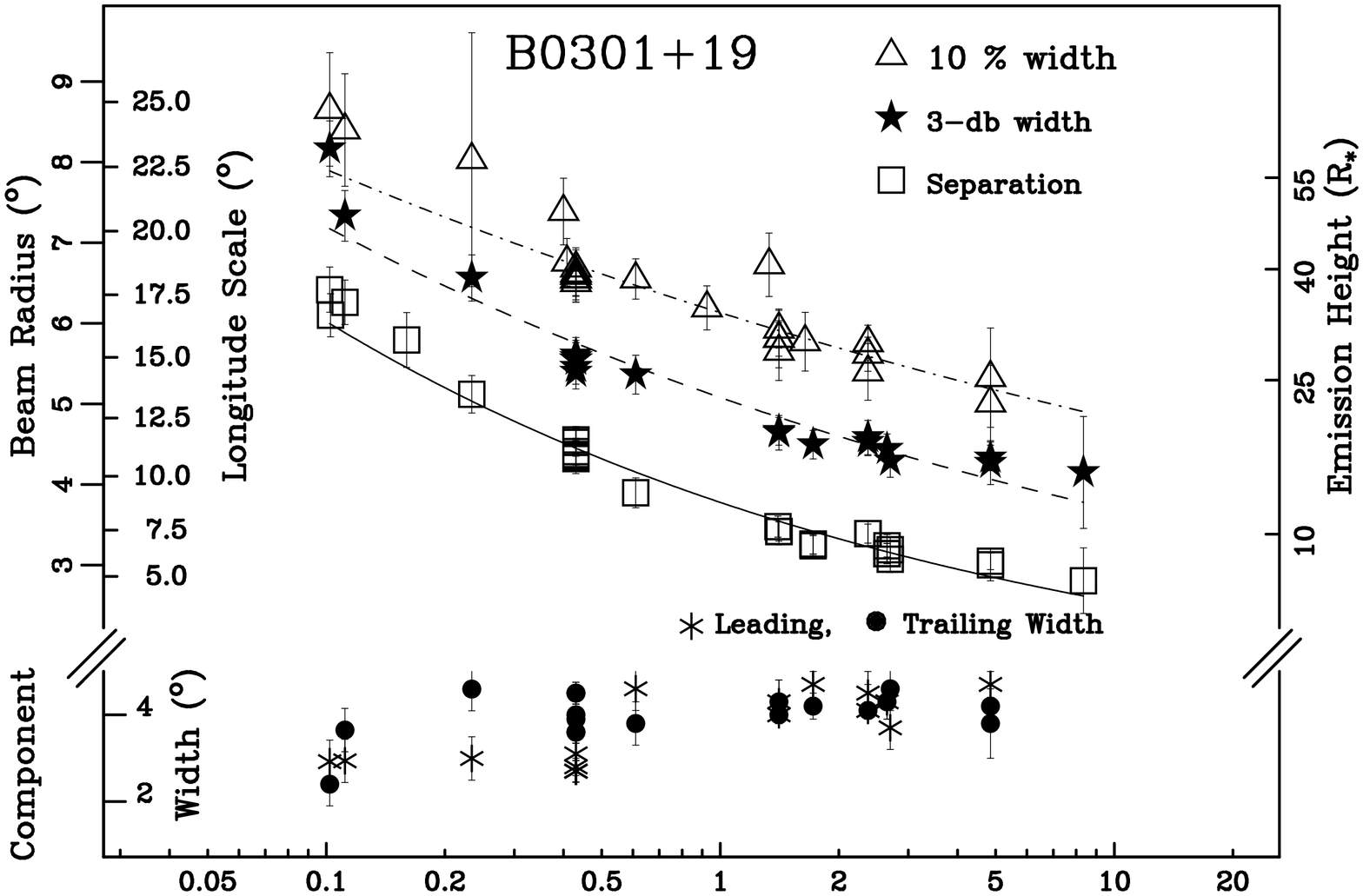,height=8cm,width=6cm,angle=-90}
\epsfig{file=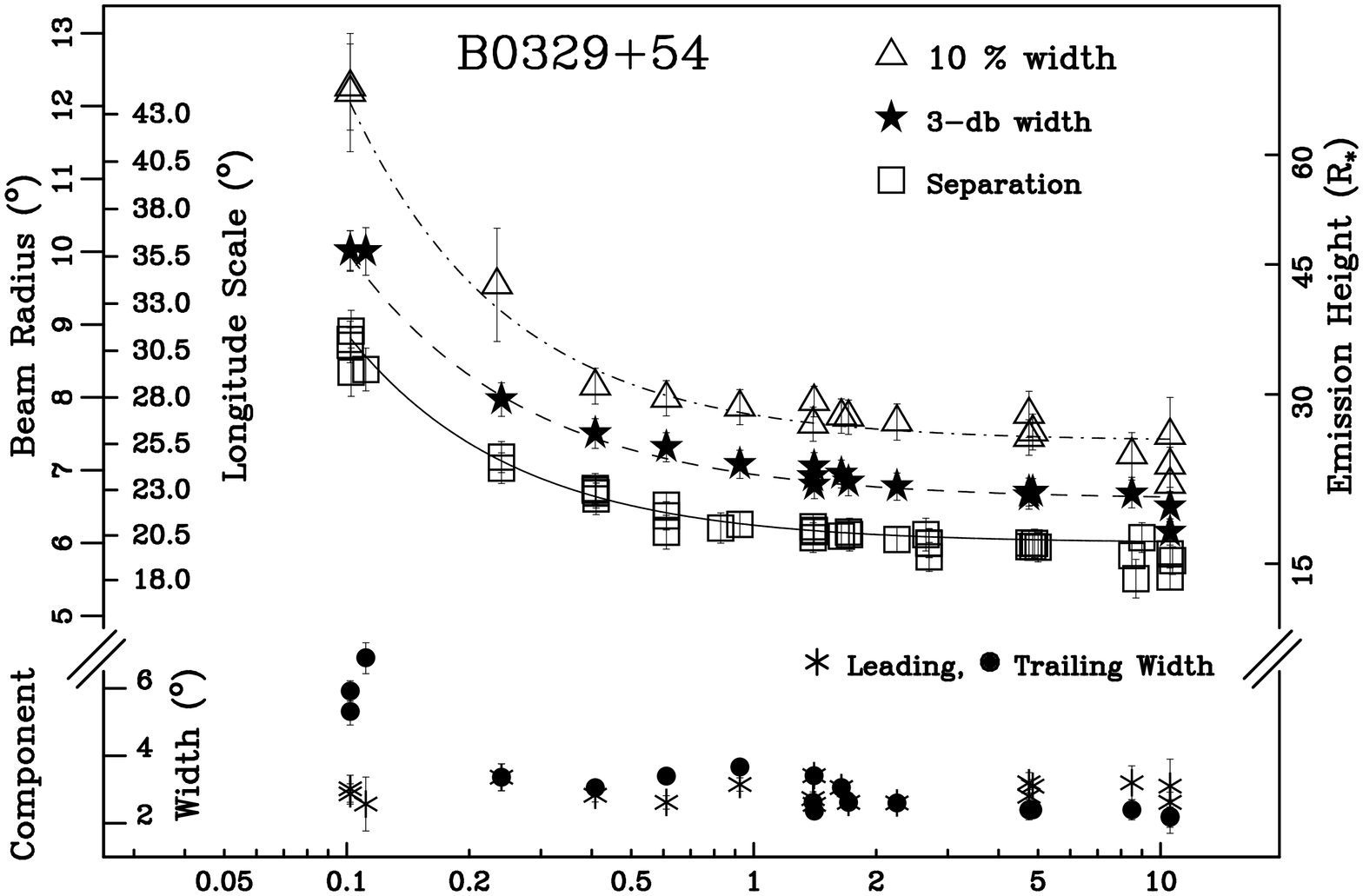,height=8cm,width=6cm,angle=-90}
\epsfig{file=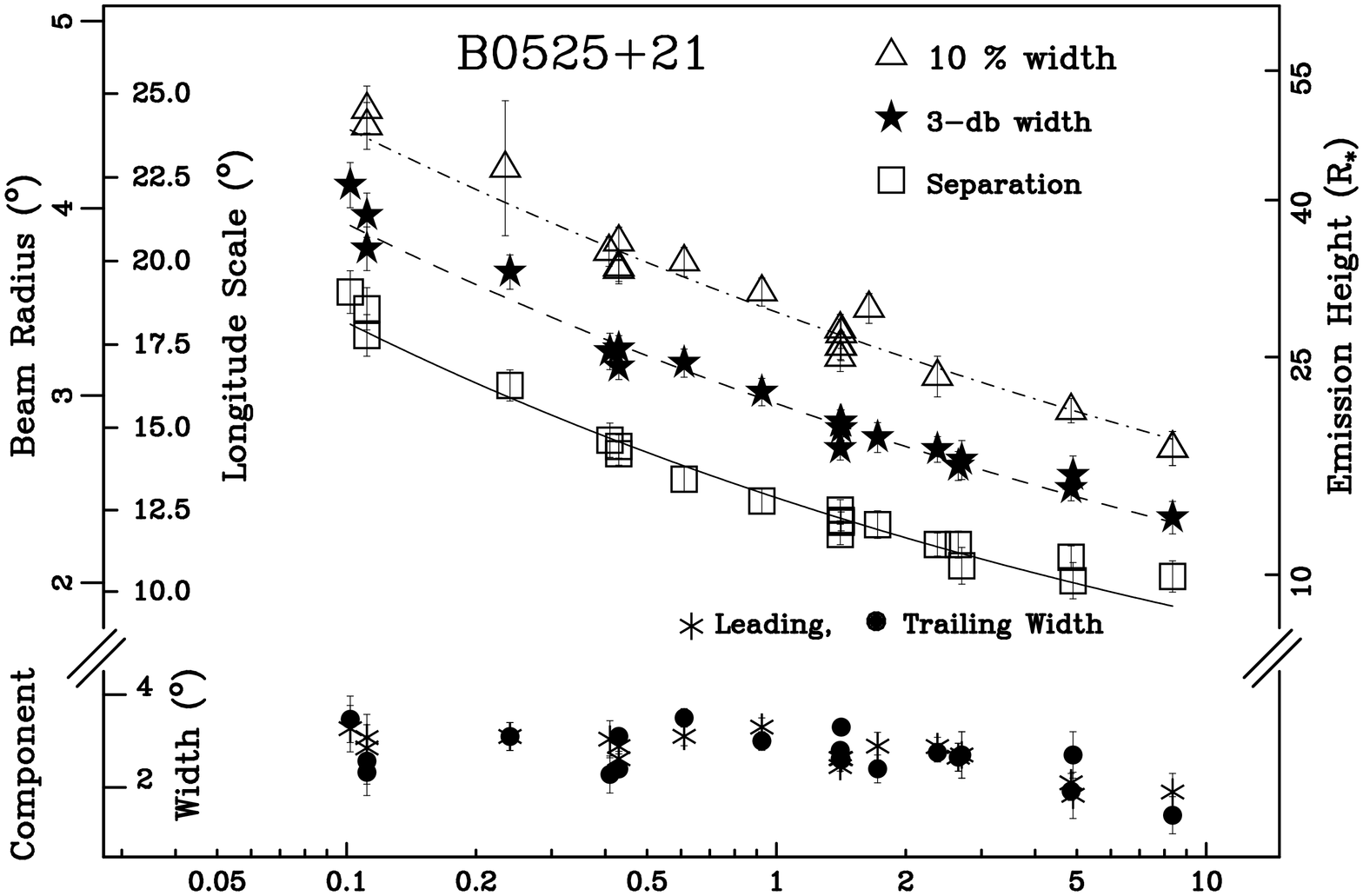,height=8cm,width=6cm,angle=-90}
\epsfig{file=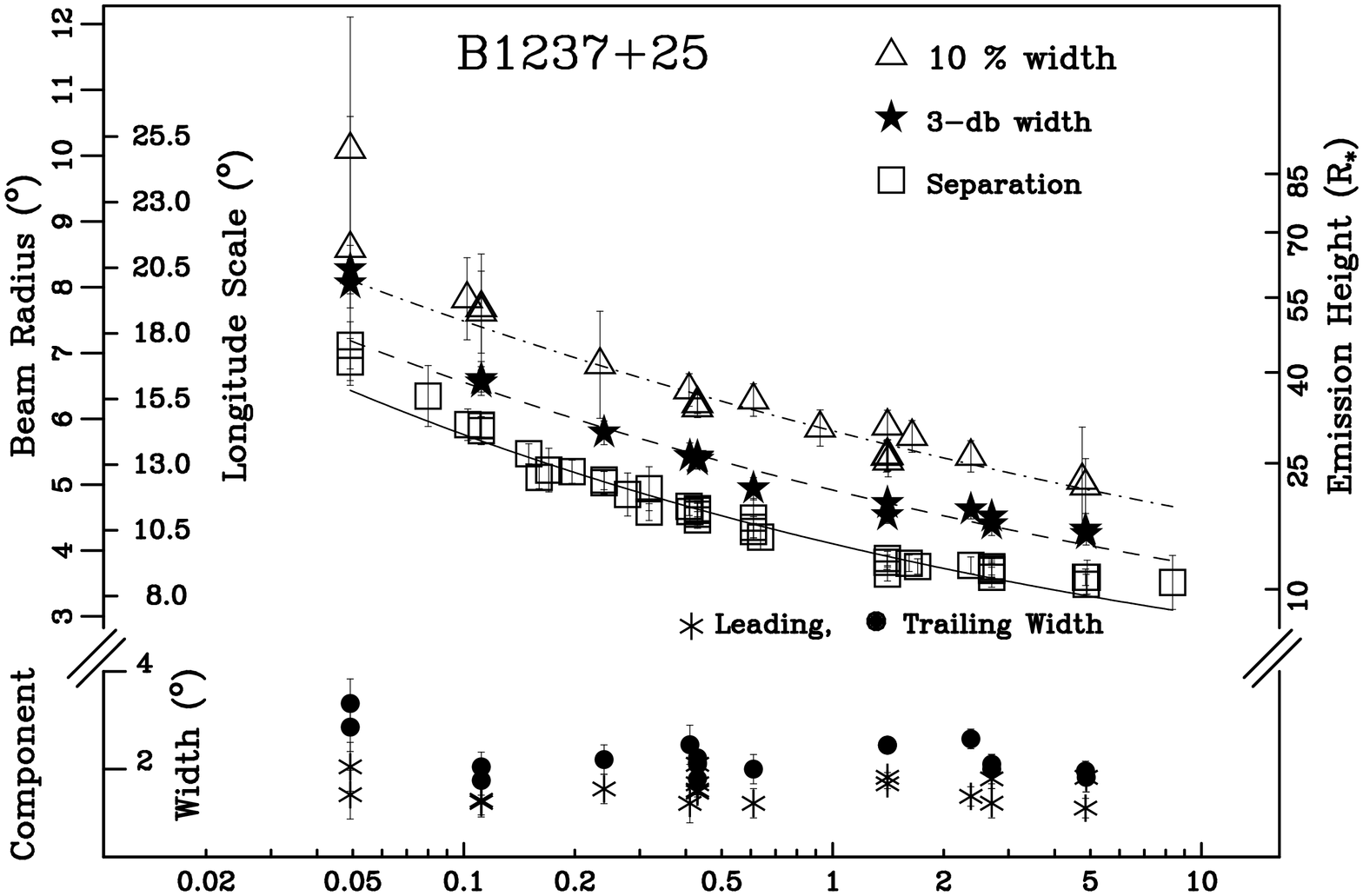,height=8cm,width=6cm,angle=-90}
\epsfig{file=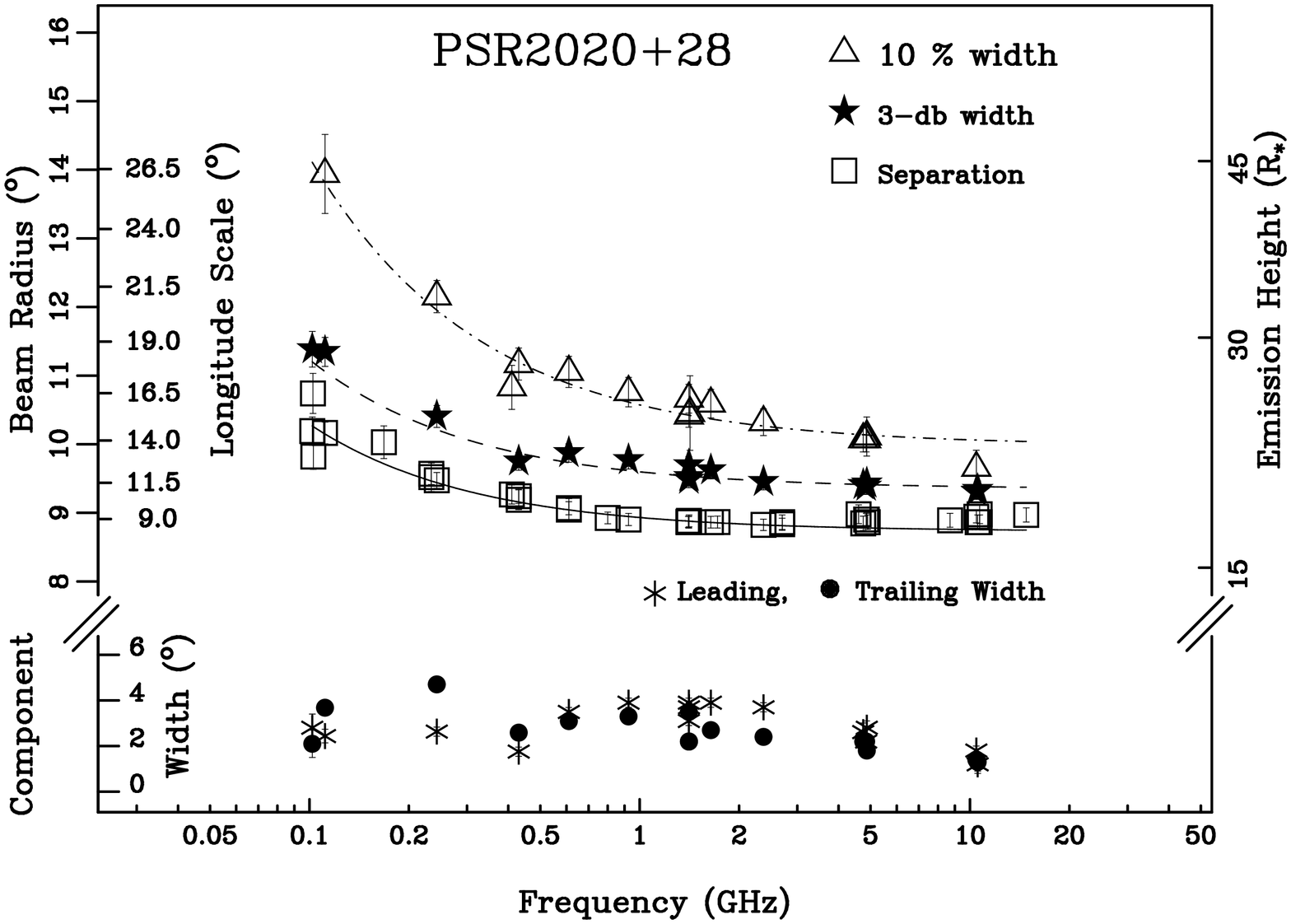,height=8cm,width=6cm,angle=-90}
\epsfig{file=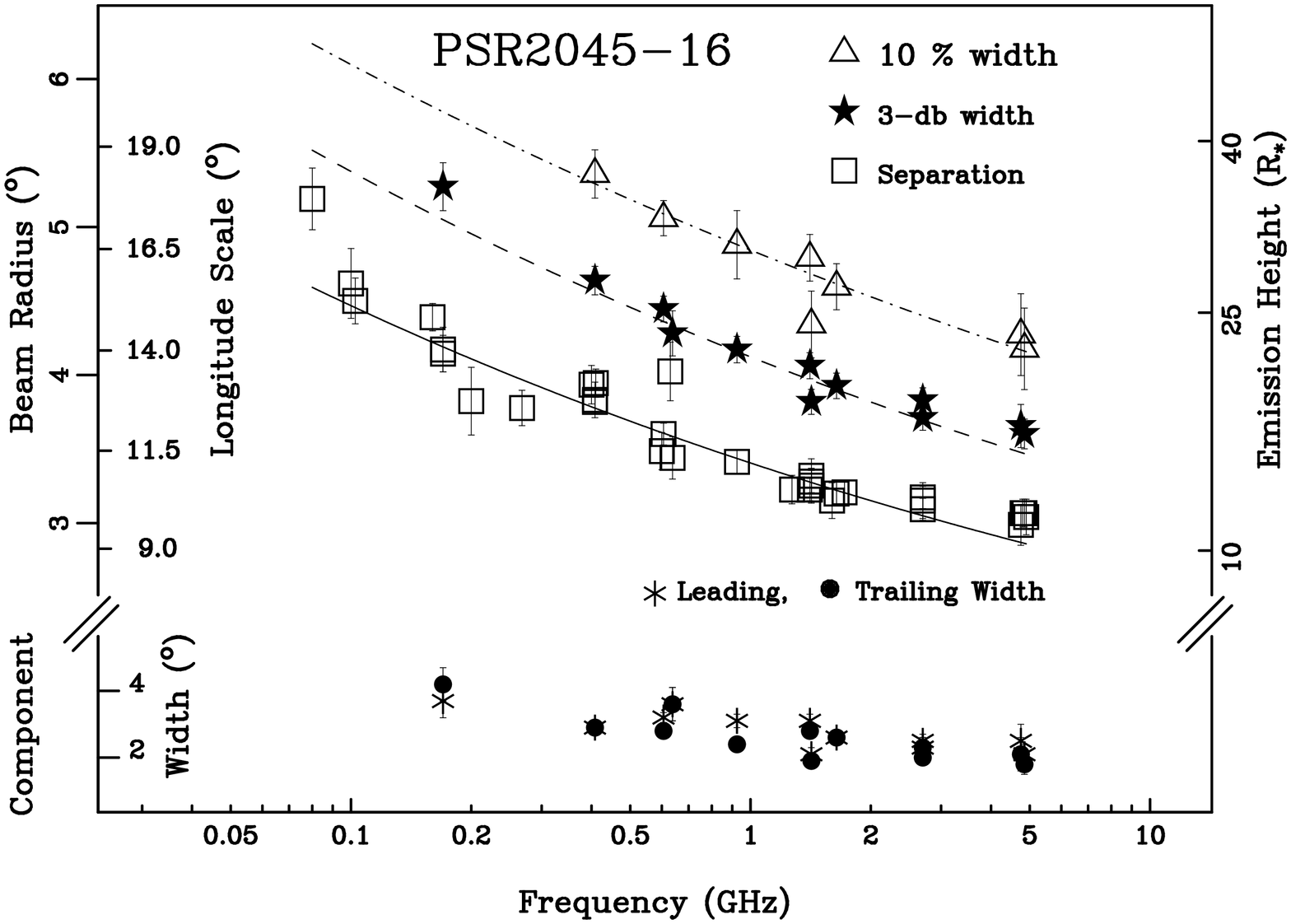,height=8cm,width=6cm,angle=-90}
\end{center}
\caption[]{Component-pair widths (lower portion), 10\% 
and 3-db profile widths, and component separations (upper 
portion) for pulsars B0301+19, B0329+54, B1133+16, B1237+25,  
B2020+28, and B2045--16 as Fig.~\ref{fig:1133wdt}.}
\label{fig:psrwdts}
\end{figure*}

The lower curves of the plots in Figure~\ref{fig:psrwdts} 
give component-width spectra for six additional stars,   
exactly as did the B1133+16 plot in Fig.~\ref{fig:1133wdt}.
Apart from the strange ``step'' in B0301+19, we see a 
relatively constant  component-width behavior.  No pulsar 
exhibits a width escalation with wavelength which at all 
mimics the corresponding profile behavior in the upper 
panels.  In order to further understand the frequency 
dependence of the component widths, we fitted power-law 
curves of the form $K f^{\zeta}$, where $f$ is in GHz. 
Table~\ref{tbl-2} then gives the resulting parameters 
over (in some cases) several ranges of frequency.  Most 
of the stars have one component---and both for B0525+21 
and B1237+25---which exhibits and an essentially constant 
width within the errors.  Only for B2045--16 do we see a 
clear spectral trend for both components---or, in other 
words, where $\zeta$ appears significant.  More often, 
the widths are ``noisy'', varying somewhat both at a 
given frequency and unsystematically over the entire 
observed band.  

We can see that even with our corrections, some of the 
lowest frequency values remain anomalously large.  It 
is possible that scattering becomes a factor, but more 
likely these lowest frequency profiles are ``smeared'' 
by unknown instrumental factors which are not completely 
correctable.  Therefore, only when future coherently 
dedispersed observations at very low frequencies become 
available can we interpret them with full confidence.  

We see no evidence for the leading or trailing component 
to be larger.  There may be some tendency for the 
components to narrow at the highest frequencies---perhaps 
resulting in the largest widths in the mid-frequency 
range.  Though this trend hardly significant for any 
particular star (perhaps for B2020+28), we find the very 
smallest widths in the extreme high frequency profiles 
in about half of the cases (see also fig. 3).  

In considering the width variations, it is also very 
possible that one or both of the two components are 
composite, with modal contributions that may have 
somewhat different temporal and spectral behaviors.  
Indeed, there is now increasing evidence that these 
conal components are comprised of modal contributions 
which have somewhat different angular dependences 
(see, for instance, Rankin \etal\ 1988; Ramachandran 
\etal\ 2002; as well as Paper VIII in this series), 
and were these to overlap increasingly in the often 
somewhat narrower high frequency components, it could 
account for the progressive average depolarization 
that characteristically occurs there ({\it e.g.} 
McKinnon 1997).  

We have entertained whether the cases of possible 
high frequency component narrowing might be associated 
with ``breaks'' in the radio-frequency spectra for 
these stars.  Such features have been identified in 
the spectra of B0301+19 (0.9 GHz), B0525+21 (1.5 GHz), B1133+16 (2.0 GHz), 
B1237+25 (0.7 GHz), B2020+28 (2.3 GHz) and B2045--16 (0.5 GHz) (Malofeev 
1994, 2001; Maron \etal\ 2000; McKinnon 1997).  For all but the second, 
the turndown frequencies seem roughly compatible with 
their width spectra, but more careful work will be 
required to fully demonstrate this association.    

{\it We conclude that these well measured component 
widths are virtually constant with frequency, apart 
from a slight tendency to narrow at high ($>$1 GHz) 
frequencies. }

\section{Conal Beam Radii}

As mentioned earlier, the conal beam radius $\rho$ is 
a far more appropriate quantity by which to assess 
RFM questions, as it is a fundamental descriptor of a 
star's emission geometry.  If we assume a {\it circular} 
emission beam, then when the magnetic latitude $\alpha$ 
and sightline impact angle $\beta$ are known, we can 
compute $\rho$ using the spherical geometry relation 
first given by Gil (1981)
\begin{equation}
\sin^2{\rho/2}=\sin\xi\sin\alpha\sin^2{\Delta\varphi/4}+\sin^2{\beta/2} ,
\label{eq2}
\end{equation} 
where $\xi$=$\alpha$+$\beta$.  $\alpha$ can be estimated 
either using the techniques of Lyne \& Manchester (1988) 
or the core-component-width method of Paper IV, wherein 
the 1-GHz core width $W_c$ is $2.45\deg P^{-1/2}/\sin\alpha$.  
Fortunately, for the three stars here with core components, 
the two methods agree very closely.  Pulsars B0301+19, 
B0525+21, B1133+16 and B2020+28 have no detectable core 
components and thus the second method cannot be used to 
determine $\alpha$. However, Paper VI found that cones 
have definite angular dimensions in terms of the polar 
cap radius, and these four pulsars all seem to have what 
were called ``outer'' cones with 1-GHz, half-power radii 
of $5.75\deg P^{-1/2}$.  So here we can work backwards, 
given a particular conal type, by then finding the value 
of $\alpha$ which squares with the characteristic 1-GHz 
value of $\rho$.\footnote{Weak arguments were used to 
classify pulsar B2020+28 as an inner-cone {\bf T} star 
in Paper VI, and viewing it as an outer conal double 
({\bf D}) object also raises problems.  Its PA traverse 
departs significantly from the usual ``S'' shape, so its 
R of $>$10 may not have the usual significance.  $\alpha$ 
must be nearly 90$\deg$ if R is as large as 8.  We used 
the somewhat smaller values in Table 1.}

Secondly, we estimate the impact angle $\beta$ as 
$\sin\beta = \sin\alpha/R$, where 
$R$ (=$|\partial PA/\partial\varphi|_{max}$) is the 
maximum sweep rate. Table~\ref{tbl-1} lists the 
values of $\alpha$, $R$ and $\beta/\rho$ (at 1 GHz) as 
taken from Paper VI.  As we do not know the sign of 
$\beta$ for any of the stars of interest here, we have 
proceeded as if $\beta$ could be of either sign---and 
the difference affects our results so little that it 
has been accommodated within the stated errors.

Assuming that $\beta$, $\rho$ and $\Delta\varphi$ are 
small and independent, the error $\sigma_{\rho}$ in 
$\rho$ can be calculated by propagating the errors   
$\sigma_\alpha$, $\sigma_{\Delta\theta}$, and $\sigma_R$ 
(given in Table~\ref{tbl-1}) as,
\begin{eqnarray}
\sigma_{\rho}^{2} = A_\alpha(1+\Delta\theta^2)\sigma_{\alpha}^2 
    + A_{\Delta\theta}\sigma_{\Delta\theta}^2 + A_R\sigma_R^2 
\label{eq3}
\end{eqnarray}
where $A_\alpha$=$(\sin\alpha\cos\alpha/4\rho)^2$, 
$A_{\Delta\theta}$=$(\Delta\theta\sin^2\alpha/4\rho)^2$, 
and $A_R$=$(\sin^2\alpha/2R^2\rho^2)^2$.  

With these considerations in mind we can now compute 
$\rho$ and $\sigma_{\rho}$ from $\alpha$, $R$ and 
$\Delta\varphi$.  We wish to fit the resulting $\rho$ 
values in order to study their frequency variation, but 
we have no {\it a priori} means of knowing how they will 
behave---or in our practical terms, what an appropriate 
fitting function might be.  However, for some pulsars 
(with small $\beta/\rho$, such as B1237+25), we know 
that the $\rho$ and $\Delta\varphi$ behavior cannot 
be very different, so that a ``recycled'' empirical 
Thorsett expression may well serve adequately for the 
$\rho$ fitting also--- 
\begin{equation} 
\rho=\rho_{\circ}+(f/f_{\circ})^a, 
\label{eq4}
\end{equation}
where $\rho_{\circ}$ and $f_{\circ}$ are constants in 
degrees and degrees-${\rm GHz}^{-a}$, respectively. 

The upper portions of Figures~\ref{fig:1133wdt} and 
\ref{fig:psrwdts} give the respective measured values, 
along with their errors, for the component separation, 
outside half-power (3-db) profile width, and the outside 
10\% width of the seven primary stars in consideration. 
Note that these primary data are plotted according to 
the slightly non-linear longitude scale on the inside, 
upper left of the diagrams.  The computed conal beam 
radii are then plotted per the linear ``Beam Radius'' 
scale on the outside, upper left of the plots.  These 
three primary measures, of course, refer to somewhat 
different positions on the conal emission beam (as well 
as incurring slightly different geometrical factors), 
and so they may or may not behave similarly when fitted.  

In order to carry out the non-linear least-square-error 
analysis, we used the Levenberg-Marquardt method as 
implemented in the {\it Numerical Recipes} routines 
(Press \etal\ 1986), fitting all three profile measures: 
separations, half-power widths, and 10\% widths.  We 
first tried unconstrained, 3-parameter fits and found 
that while the Thorsett expression fitted the profile 
measurements adequately---that is, with $\chi^2$ values 
near unity---the parameters were all highly correlated.  
Table~\ref{tbl-3} summarizes the fitted parameters, 
normalized correlation coefficients, reduced $\chi^2$ 
values, and number of fitted points $M$, for some of 
these fits.  The correlations between $\rho_{\circ}$ and 
$f_{\circ}$ ran between +0.74 and +0.93, those between 
$f_{\circ}$ and $a$ around --0.98, and the third one 
also negative and intermediate in magnitude. 

In an effort to provide some physical constraint on the 
fitting function, we attempted to interpret $\rho_{\circ}$, 
the infinite-frequency value of $\rho$, in terms of the 
field direction at polar cap edge $\rho_{\rm pc} = 1.23\deg P^{-1/2}$.  
This $\rho_{\rm pc}$ is associated with the ``last open 
field lines'' and may or may not represent the radius 
of the ``active'' region on the stellar surface, but it 
seemed meaningful to ask whether such a small asymptotic 
value could be compatible with the overall frequency 
dependence of the conal beam radius.  Moreover, we do 
not know which profile measure, if any, might correspond 
to this outer boundary of the polar cap.  Therefore, we 
explored the behavior of all three.

It was thus surprising to find that upon fixing $\rho_{\circ}$ 
as $\rho_{\rm pc}$, very adequate fits were obtained for 
four stars, B0301+19, B0525+21, B1237+25 and B2045--16 
(hereafter Group A).  The results are given at or near 
the top of each section in Table~\ref{tbl-3}---with the 
fixed $\rho_{\circ}$ values shown in boldface---and the 
three respective fits for each pulsar are hardly better 
in one case or the other as judged by the $\chi^2$ values.  
It thus seems that the A-group stars are insensitive to 
assumptions about what point on their profiles corresponds 
to the ``last open field lines''.  The A-group $f_{\circ}$ 
values are well beyond the highest frequency observations, 
reflecting the fact that it is only here that the second 
term in eq.(\ref{eq4}) declines to a value of 1$\deg$.  
Qualitatively, we see this behavior clearly in 
Figs.~\ref{fig:1133wdt} and \ref{fig:psrwdts} 
as the A-group measures continue to decrease at the highest 
frequencies (in marked contrast to the remaining stars,
which approach an asymptotic width at high frequency).

Then, we asked whether reasonable fits for the A-group 
stars could be obtained if $\rho_{\circ}$ was fixed as 
above {\it and} the index $a$ was set to the (plasma 
frequency associated) value of $-1/3$---and the answer 
was negative.  We give these fitting results only in 
terms of the component separation, and the corresponding 
entries in Table~\ref{tbl-3} show that the $\chi^2$ 
values fall far above unity.  Finally, we asked whether 
fixing $a$ at $-1/3$ was compatible with the observations 
if the constraint on $\rho_{\circ}$ was relaxed, and 
our results for the separation measure appear just under 
the foregoing ones.  Again, to our surprise, we found 
that these fits were very acceptable (slightly better 
than those obtained by constraining $\rho_{\circ}$), 
with values of $\rho_{\circ}$ well larger than 
$\rho_{\rm pc}$ by factors of between $3/2$ and 3.  

However a similar exercise for PSRs B0329+54, B1133+16 
and B2020+28 (hereafter Group B) shows a very different 
behavior. The first group of (separation) results quoted 
for these stars in Table~\ref{tbl-3} represents the 
unconstrained fits. For these stars, the parameters are 
slightly better determined as $\rho_{\circ}$ and $f_{\circ}$ 
are only about 80\% correlated.  This is so because 
here the $\rho$ values saturate rather rapidly with 
frequency---{\it i.e.} at 300 MHz or so---as can be seen 
by the very different $f_{\circ}$ values for the B-group 
stars.  Again, apart from the half-power width fit for 
B1133+16 (where $\chi^2$$\sim$4), the profile measure has 
very little effect on the fitting quality.

As might be expected from the qualitative behavior, no 
acceptable fits could be obtained for the B-group stars 
when $\rho_{\circ}$ was constrained to $\rho_{\rm pc}$.  
When the index $a$ was fixed at $-1/3$ as above though,
Table~\ref{tbl-3} shows that all the $\chi^2$s fall just 
above the acceptable range.  Moreover, the $\rho_{\circ}$ 
values are all about 3 times greater than $\rho_{\rm pc}$, 
just as for the A-group stars.  The $f_{\circ}$ values 
remain low, 2--10 times those for the free fits, and 
representing frequencies within the range observed.  

{\it Summarizing then, conal beam radii computed from the 
three profile measures are well fitted by eq.(\ref{eq4}) 
with unconstrained parameters, though with very high 
correlations.  Only for the A group, however, do 
acceptable fits obtain when $\rho_{\circ}$ is fixed at 
$\rho_{\rm pc}$ or $a$ to --1/3.}

\section{Do Inner Cones Exhibit RFM?}

\begin{figure}
\begin{center}
\epsfig{file=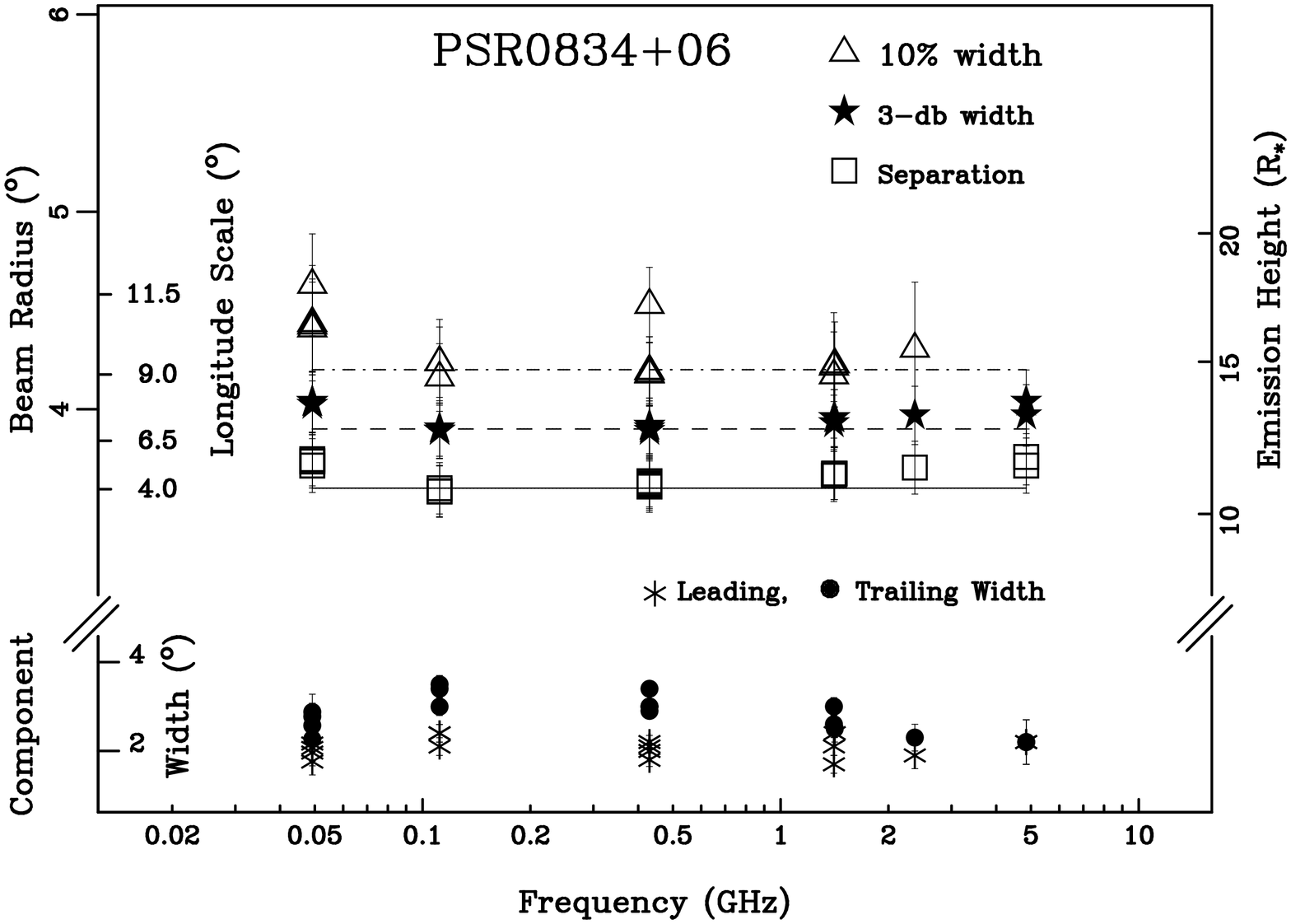,height=8.0cm,angle=-90}
\epsfig{file=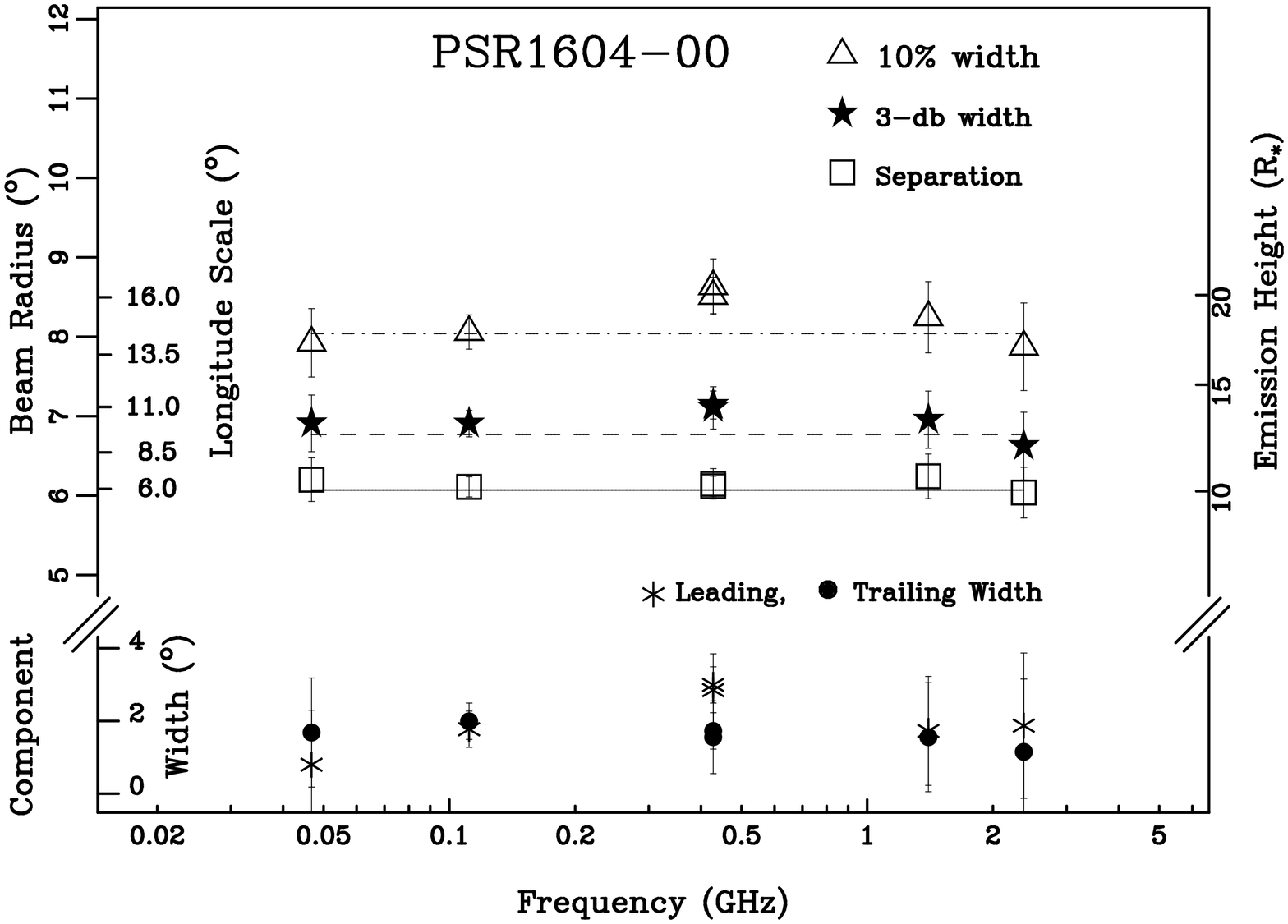,height=8.0cm,angle=-90}
\epsfig{file=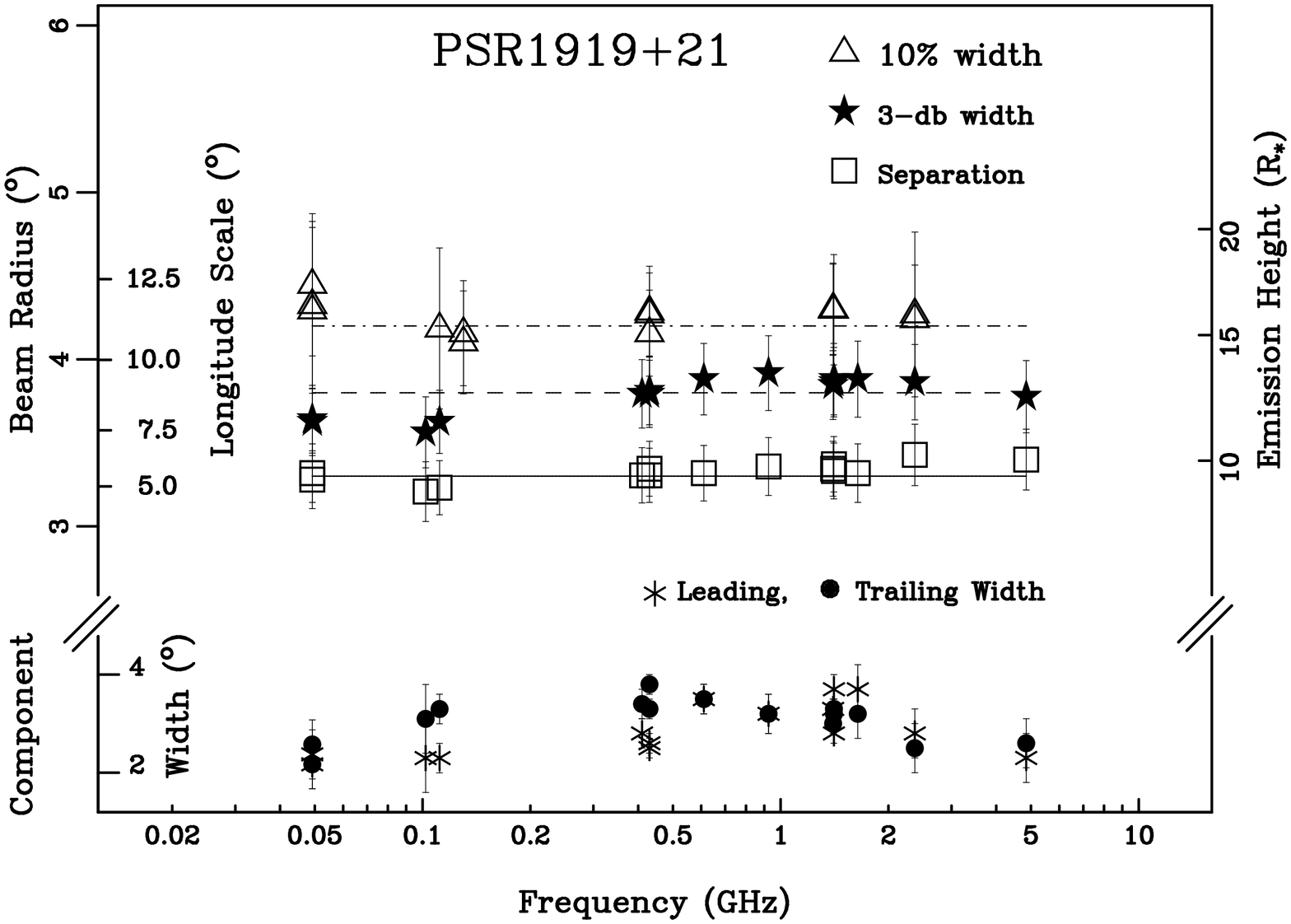,height=8.0cm,angle=-90}
\end{center}
\caption[]{Component-pair widths (lower portion), 
10\% and 3-db profile widths, and component 
separations (upper portion) for pulsars B0834+06, 
B1604--00 and B1919+21 as Fig.~\ref{fig:1133wdt}.  
Note that these stars exhibit almost no change 
in their component separations and very little 
in their profile widths.}
\label{fig:psraddl}
\end{figure}

Figure~\ref{fig:psraddl} gives similar displays for three 
additional pulsars, B0834+06, B1604--00 and B1919+21.  All 
exhibit a very different behavior---a near constancy of 
their profile dimensions.  Sieber \etal's (1975) study 
found, for B0834+06 and B1919+21, a slight width escalation 
both above and below about 200 MHz, trends which can be 
seen in our plots as well.  Of course, this behavior does 
not permit fitting eq.(\ref{eq4}) to the profile measures, 
so instead we plot lines showing the best-fitting constant 
values.  Overall, the separation is virtually fixed for 
these objects within their errors.  The profile width 
measures show somewhat more variation, in part reflecting 
changes in the component widths, which tend to be largest 
at intermediate frequencies.   The one-parameter fitting 
results for these stars are given in Table~\ref{tbl-3} 
as Group C.

The ostensible complete lack of RFM in these Group C 
stars is so striking in contrast to the foregoing A- 
and B-group pulsars that it demands further inquiry.  
A hint comes from B1604--00, whose triple {\bf T} 
profile's core-component width can be determined 
accurately (Rankin 1986), and modeling its emission 
geometry (Paper VI) indicates that its conal component 
pair represents an inner cone.  The star's three primary 
components overlap, but they can be segregated either 
polarimetrically or by Gaussian fitting.  Here the high 
quality Arecibo profiles of Hankins \etal\ (2002) have 
been especially useful.  
  
No core component has been identified in the other 
two group-C pulsars, so it is not possible to be 
as definite about their geometry; however, the overall 
structure of both could well be of the inner-cone type.  
Pulsar B0834+06 was analyzed as having an inner cone in Paper 
VI; whereas the complexity of B1919+21's emission ({\it 
e.g.}, Pr\'osynski \& Wolszczan 1986) then tended to 
suggest a double-cone structure---in turn then implying 
an outer cone---but with recent evidence for one smaller 
``further-in'' cone (Gil \etal\ 1993; Rankin \& Rathnasree 
1997; Gangadhara \& Gupta 2001; hereafter G\&G), B1919+21's 
conal structure could still be of the inner type.\footnote{Additionally, 
the three pulsars share a further unusual characteristic:  
their profiles exhibit low level emission well outside 
the two main conal features, particularly on their 
leading edges.  In B0834+06 this can be seen very clearly 
in the recent work of Weisberg \etal\ (1999), but other 
observations also show a ``ramp'' or weak leading-edge 
feature ({\it e.g.} Kardashev \etal\ 1986).  In B1604--00's 
profile the early emission is very prominent.  Weisberg 
\etal's profile clearly shows the inflections corresponding 
to the core and two conal features as well as marked 
emission well before them.  Finally, we see a very similar 
behavior in B1919+21; the profiles in Hankins \etal\ (2002), 
Gould \& Lyne (1998), and Weisberg \etal\ (1999) all 
show a leading-edge feature which appears to be 
strongest at very low and high frequency---and some 
suggest similarly enhanced emission on the trailing 
edge also.  In fact, in the 4.85-GHz profile in von 
Hoensbroech's (1999) thesis, this emission appears as 
a distinct preceding component.                        \\ 
So, in short, if the primary conal component pairs in 
these stars are inner cones, then perhaps this highly 
unusual emission on the fringes of their profiles 
should be interpreted as evidence for weak outer-cone 
emission.  It is worth noting that these three pulsars 
are some of the strongest and least dispersed in the 
sky, so we have a much greater opportunity to obtain 
sensitive profiles wherein low level emission could 
be detected over a wider frequency range (including 
on an individual-pulse basis).  Conversely, the Group 
A and B stars are at least equally bright and little 
dispersed, and we see nothing of this weak outlying 
emission in these stars.                           \\ 
All three group-C stars deserve major new observational 
and analytical attention:  Two are Cambridge pulsars 
which have had very little recent study.  All would 
benefit from the single-pulse polarimetry techniques 
developed since the 1990s, carried out over a wide 
frequency range.}

Interestingly, the width spectra of these three 
stars---with demonstrated or probable inner-cone 
configurations---have very much in common with what is 
known about the inner cone behavior in the double-cone, 
five-component {\bf M} pulsars that were studied in 
Paper VI.  Unfortunately, only three, B1237+25, 
B1451--68 and B1857--26, have been studied sufficiently 
well over a wide enough frequency range to give even 
a qualitative indication of inner-cone separation 
spectra.  These data are tabulated in Paper VI, 
Table 3, and for the first two stars it can be seen 
that the inner cone, half-power pair widths are either 
nearly constant or much more slowly varying than the 
corresponding outer conal pair.  For B1857$-$26, the 
existing observations provide only the qualitative 
impression that the inner cone dimensions remain 
relatively constant.\footnote{Again, much more work 
is needed to elucidate not only the behavior of 
these three ``exemplars'' but also the many other 
and/or more recently discovered {\bf M} pulsars.}                          

One additional means of attempting to confirm this 
striking lack of RFM in inner conal component pairs 
is that of examining the behavior of other stars, 
like B1604--00, with inner-cone triple {\bf T} profiles.  
We found in Paper VI that about half the {\bf T} stars 
had inner cones (23 of 40), and recent observations---particularly 
by Gould \& Lyne (1998; hereafter G\&L)---now provide 
good quality profiles over a wide enough band (1642 
to 234 MHz) to see if their conal configurations 
change very much.  A full analysis of this question 
goes beyond the scope of this paper, because most of 
the inner cone {\bf T} stars will require fitting 
to accurately locate their components.  However, we 
can get a useful qualitative impression.  Of the 22 
pulsars apart from B1604--00 (and using G\&L, Hankins 
\etal\ 2002, and the sources in Paper VI), 13 stars, 
B0149--16, B0450+55, B0450--18, B0919+06, B1221--63, 
B1508+55, B1541+09, B1702--19{\it m}, B1706--16, B1727--47, 
B1747--46, B1818--04, and B1839+09, have profiles in 
which the inner-conal pair spacing changes hardly at 
all.  Further, none of the 9 others provides a contrary 
example (most have not been observed at low enough 
frequency, are scattered, or are simply too weak to 
bear on the question).  

{\it The conclusion then seems inescapable: the conal 
beams comprised by inner cones exhibit virtually no 
RFM---that is, inner conal component pairs show little 
or no change in their longitude separation---in turn 
implying that the inner conal beams which produce them 
have radii that vary little or hardly at all with 
frequency.}  The RFM behavior of the outer cones of 
Groups A and B and the inner cones of Group C (and 
the other stars mentioned above) could hardly be more 
different.  The former exhibit the rather dramatic 
RFM which has attracted the attention of many 
investigators; whereas, the ostensibly similar inner 
cones (differing only in their distinct characteristic 
angular size relative to that of the polar cap 
$\rho_{\rm pc}$) show no discernible RFM at all.  

\section{Reconciling the Conal Beam Measures}

Let us now compare the results of the above analysis with 
the earlier published work on conal beam dimensions.  We
mentioned above the conclusion (Paper VI) that most conal 
beams have two characteristic outside, half-power radii, 
$\rho_{\rm inner[outer]} = 4.33\deg[5.75\deg] P^{-1/2}$ 
at 1 GHz.  In addition, there are further results from Gil 
\etal\ (1993), Kramer \etal\ (1994), Rankin \& Rathnasree 
(1997), Mitra \& Deshpande (1999), and Gangadhara \& Gupta 
(2001). In analyzing their profile-dimension information, 
these authors used different measures, and now, using our 
analysis tools, we can find a relationship between them.  

We have seen above that the various measures fit the 
respective profiles with somewhat different parameters.  
However, the largest part of this difference is in the 
very different spectral behavior of the 3-db and 10\% 
component widths.  As these widths are fairly constant, 
very similar values for $f_{\circ}$ and $a$ should be 
obtained in fitting any of the three profile measures; 
the difference will appear almost entirely in $\rho_{\circ}$.  
Therefore, we proceed here by imposing values of $f_{\circ}$ 
and $a$, obtained from fitting the separations, on the 
profile-width fits in order to study their relationship.

Specifically, for the A-group pulsars, we imposed 
$f_{\circ}$ and $a$ values, obtained from the 
$\rho_{\circ}$=$\rho_{\rm pc}$-constrained, component-separation 
fits in order to obtain the best fitting values of 
$\rho_{\circ}$, $\rho_{\rm cs}$=$\rho_{\rm pc}$, $\rho_{\rm
3db}$, and $\rho_{\rm 10\%}$.  
We computed $\rho_{\rm 3db}-\rho_{\rm cs}$ and
$\rho_{\rm 10\%}-\rho_{\rm cs}$ for each star, and 
found that the ratios were in reasonable agreement, 
yielding an average value of $0.51\pm0.15$. Therefore, 
the 10\% point on the conal beam seems to fall twice as 
far from the peak as from the half-power point.  A 
similar ratio can be gleaned from the results of Gil 
\etal\ (1993), whose ensemble 3-db and 10\% points fall 
at 5.52$\deg$$\pm$0.5$\deg$ and 6.33$\deg$$\pm$0.5$\deg$, 
respectively, in relation to a peak value at 4.6$\deg$.  
This dependence is not far from that of a Gaussian curve, 
where the half-power point falls at 0.55 that of the 
10\% point.

A similar procedure was carried out on the Group B and 
C stars (without the $\rho_{\circ}$=$\rho_{\rm pc}$ constraint).  
For the B group, the component-separation fit values can 
be imposed on the other two measures only if some high 
and low frequency values are omitted (because of changes 
in the component widths), and for the C group only an 
overall $\rho_{\circ}$ can be defined.  The average 
values of the ratios was thus 0.42$\pm$0.1 and 0.45$\pm$0.1 
for Group B and C, respectively---slightly lower perhaps 
than for the Gaussian curve.  

{\it Overall then, we conclude that conal beams have a 
roughly Gaussian dependence in $\rho$.}

\section{Emission Height vs. Frequency}

We now consider the implied heights of emission, 
assuming first a dipole structure for the stellar 
magnetic field, and second that the emission at a 
given frequency arises from the same altitude 
above the center of the 10-km neutron star.  The 
radio emission height $h$ can thus be computed as 
\begin{equation}
h= 10 P (\rho/1.23\deg)^2~  \rm{km.}
\label{eq6}
\end{equation}
Then, since $\rho$ is a function of frequency as 
given by the neo-Thorsett relation eq.(~\ref{eq4}), 
the dependence of $h$ on frequency---RFM mapping---can 
be studied  directly.  The upper portions of all 
the foregoing figures show this computation of $h$ 
on their outside, right-hand scales in units of 
the (putative 10-km) stellar radius $R_*$.  Because our 
analysis used values of $\alpha$ and $\beta$ 
estimated by one of us (Paper VI)---which in turn 
used the convention of associating the emission 
with the ``last open field lines''---the 1-GHz, 
3-db-width heights for the group A and B stars 
(all shown or thought to be outer cones) in 
these figures are all just over 200 km.  The 
heights corresponding to the separation and 10\% 
widths are consequently smaller and larger, 
respectively.  

It is certainly possible to relate the fitting data 
in terms of $\rho$ [eq.(\ref{eq4})] to the height 
expression above.  In that $h \propto \rho^2$, the  
fitting function must be squared, and 
the uncertain effect of the cross term complicates 
interpretation. A more transparent approach is that 
of fitting the frequency dependence of $h$ directly.  
In that we have no {\it a priori} knowledge of $h$=$h(f)$, 
we want to assess this approach to ensure that our 
results are meaningful.  Again, we try a version of 
Thorsett's relation
\begin{equation}
h = h_{\circ} + (f/f_h)^{b}  ,
\label{eq7}
\end{equation}
where $h_{\circ}$ and $f_{h}$ are constants in 
units of km and GHz-km$^{-1/b}$, respectively.

In order to fit eq.(\ref{eq7}) we used the separation 
measure, as use of the others would lead to similar 
results, while raising the complications discussed 
above.  The results of the fits are summarized in 
Table \ref{tbl-4}.  The fitting procedure was 
identical to that for eq.(\ref{eq4}) above.  Once 
again, the unconstrained, 3-parameter fits are 
presented first for both the group A and B stars, and 
both give uniformly good $\chi^2$ values.

In an effort to understand our earlier fitting 
results for $\rho$=$\rho(f)$, we attempted to test 
the situation where $h_{\circ}$ is fixed to $R_*$ 
(corresponding to $\rho$=$\rho_{\rm pc}$) and the 
index $b$=$2a$---obtaining unsatisfactory fits (not 
shown) for both the A and B groups ($\chi^2$$\sim$200 
for the latter).  However, the middle set of A-group 
results shows that decent fits do obtain with 
$h_{\circ}$ set to $R_*$, but with $b$ values somewhat 
different than $2a$.  

Then we asked if reasonable fits are obtained if 
$h_{\circ}$ is constrained to $R_*$ and $b$ to the 
plasma-frequency-associated value of $-2/3$, and 
again the answer was negative for both groups.  
However, if $h_{\circ}$ remains unconstrained, very 
reasonable fits are obtained for all the A-group stars 
as reported in Table~\ref{tbl-4}, though the resulting 
$h_{\circ}$ values are 5--10 times larger than $R_*$.  
In comparing these results with the heights inferred 
from the corresponding $\rho_{\circ}$ values in 
Table~\ref{tbl-3} where $a$=--1/3, we find that the 
latter are only slightly smaller---suggesting that 
the overall effect of the cross-term is small.  

Finally, we turn to the B group and carry out the 
same fitting procedure, first freeing all three 
parameters and then fixing only $b$ at $-2/3$.  
Surprisingly, reasonable fits were obtained in both 
cases, the former having only slightly lower values 
of $\chi^2$ as shown in Table~\ref{tbl-4}.  Recall 
that when group B fits were made for $\rho$ in 
Table~\ref{tbl-3}, holding $a$=$-1/3$, the results 
were were rather poor.  

{\it It is indeed remarkable that by fitting $h$=$h(f)$ 
directly using eq.(\ref{eq7}), both the A- and B-group 
stars exhibit a consistent behaviour with the index 
$b$=$-2/3$ and with $f_{\circ}$ values falling in a 
relatively narrow range between about 50 and 150 km.  
These fits suggest that only an index value near 
$-2/3$ could accommodate the different characteristics 
of the A and B groups.}

\section{More Physical Height Estimates}

The height estimates made by one of us in Paper VI 
are in need of revision.  The convention there of 
associating the outside 3-db points of the component 
pairs with the full polar cap radius (or the ``last 
open field lines'') is not very physical, first 
because the electric field ${\bf E}$ is expected 
to vanish along this boundary and second because we 
are interested in the effective height of the 
emission region, not its lower edge. 

A better convention, we now believe, is that of 
associating the emission with an active annulus on 
the polar cap. This is furthermore compatible with 
an $\bf{E}$$\times$$\bf{B}$ origin of subbeam 
circulation in B0943+10 (Deshpande \& Rankin 1999, 
2001; Asgekar \& Deshpande 2001; Rankin, 
Suleymanova \& Deshpande 2002) and is implicit in 
Ruderman \& Sutherland's (1975; hereafter R\&S) 
theory.  It also follows from the recent 
aberration/retardation (hereafter A/R) delay 
analyses of B0329+54 by Malov \& Suleymanova 
(1998; hereafter M\&S) and G\&G.  To this end we 
can modify eq.(\ref{eq6}) to include the parameter 
$s$, giving the fractional radius of the ``active'' 
annulus on the polar cap; see eq.(\ref{eq8}) below 
(also Kijak \& Gil 1997, 1998).  The A/R studies 
argue that $s$$\approx$0.5, implying that the $h$ 
values given by eq.(\ref{eq6}) are low by factors 
as large as 4.

\begin{equation}
h= 10 P (\rho/s 1.23\deg)^2~\rm{km.}
\label{eq8}
\end{equation}

Specifically, for B0329+54, M\&S used 0.06--10.7-GHz 
observations and G\&G 325- and 610-MHz observations 
to estimate their emission heights. Measuring 
the conal component spacings relative to the 
central (core) component (which is thought to be 
emitted at low altitude along the magnetic axis), 
and interpreting them as due to A/R delay, 
appropriate emission heights could be computed.  
Further, G\&G reported that four pairs of conal 
components could be identified in the star's pulse 
sequences, which in turn exhibited progressive and
 symmetrical amounts of A/R delay.  Our outer-cone 
analysis refers to their cone 3; whereas, M\&S 
considered only this outer cone.  The resulting 
conal emission heights at 103/111, 325 and 610 MHz 
are then some 945$\pm$55, 770$\pm$110, and 600$\pm$180 
km, respectively, with respect to the core emission 
height.  Using our separation measure, we find conal 
emission heights of $345\pm188$, $229\pm61$ and 
$199\pm32$ km, respectively.  Comparing these values 
using eq.(\ref{eq8}), we find an average $s$ of 
0.56$\pm$0.1; however, were the core emitted at a 
significant height, the former values would be 
increased by this amount and the resulting $s$ 
values consequently decreased.      

Almost a decade earlier, Blaskiewicz, Cordes, and 
Wasserman (1991; hereafter BCW) developed a method 
of estimating emission heights using the slight 
aberrational shift of the PA traverse relative to 
the profile center; and they then applied their 
technique to a group of stars for which they had 
obtained exceedingly well measured polarization 
profiles---and a few of their pulsars are common 
to our sample.  Then, more recently, von Hoensbroech 
\& Xilouris (1997b) applied the same technique to 
another group of stars at somewhat different radio 
frequencies, and they also considered several 
stars of interest here.  

Using the above results, three height values for 
B0301+19 were available at various frequencies, 
which could be compared with our (separation) height 
relation [Table 4; eq.(\ref{eq7})] via eq.(\ref{eq8}) 
to determine the parameter $s$.  Given our picture 
that a single active annulus on the stellar polar 
cap produces a cone of active field lines (which 
then radiate at different heights), it follows that 
$s$ should be a (frequency-independent) constant 
for a given pulsar. The measured B0301+19 $s$ values 
are 0.96$\pm$0.3, 1.3$\pm$0.3, and 0.23$\pm$0.2 at 
410, 1410 and 4850 MHz, respectively---giving a 
weighted average of 0.66$\pm$0.15.  The five available 
values for B0525+21, 0.88$\pm$0.3, 0.66$\pm$0.2, 
0.76$\pm$0.1,0.41$\pm$0.15 and 0.53$\pm$0.12 at 430, 
1410, 1418, 1710 and 4850 MHz, respectively, yield 
a weighted-average $s$ of 0.63$\pm$0.06. For B1133+16, 
five further values, 0.85$\pm$0.3, 0.61$\pm$0.1, 
0.82$\pm$0.3, 0.72$\pm$0.1 and 0.62$\pm$0.1 at the 
same frequencies, give  0.61$\pm$0.06.  For B2045--16, 
only a 4850-MHz exists, 0.54$\pm$0.1.  While there is 
considerable variation in the various individual $s$ 
estimates, the weighted-average (outer cone) $s$ 
values all fall within a narrow range between about 
1/2 and 2/3.

We can now adjust the nominal 1-GHz, outer-cone, 
emission height of Paper VI in light of the above 
results, first by using the component-separation 
measure and second an appropriate $s$ value.  The 
effect of the first issue can be estimated by 
comparing the 3-db and separation height values 
at 1 GHz for A-group pulsars B0525+21, B1237+25 
and B2045--16, and we find that the latter value is 
some 152$\pm$7 km (rather than the about 220-km 
value based on 3-db widths).  The actual physical 
height is then $s^{-2}$ times this value---or some 
610$\pm$25 km if $s$ is 0.5 or some 345$\pm$17 km 
if $s$ is 0.67.  

These outer-cone values are to be compared
with the inner-cone emission heights of the group 
C pulsars, which are both smaller and virtually 
invariant.  Here, comparing the 3-db versus the 
separation widths in Table 3, we find an average 
ratio of some 0.89, which makes the nominal 
inner-cone emission height (using the separation 
measure) some 104$\pm$6 km [as opposed to the 
Paper VI (3-db width) values of 130 km].  This 
nominal inner-cone height, when adjusted as above 
by $s^{-2}$, gives physical emission-height values 
between 415 and 235 km.  

Finally, Wright (2002) has suggested that the 
outer-cone emission region is associated with the 
last open field lines, whereas the inner-cone 
region is along the ``null'' surface (Goldreich 
\& Julian 1969; R\&S) at $s$=$(2/3)^{3/4}$=0.74.  
If so, then the above arguments would suggest a 
physical 1-GHz, outer-cone emission height of 
152$\pm$7 km as well as an inner-cone height of 
104$\pm$6 km, scaled by $s^{-2}$, which is 
191$\pm$11 km.  Note that this places in the 
1-GHz inner-cone emission at a greater height 
than the 1-GHz outer-cone radiation, though RFM 
would reverse the situation at most lower 
frequencies.  

{\it Thus, we conclude that the 1-GHz emission heights 
$\rho_{\rm inner[outer]}$ are 2--3 times larger than 
the respective 130[220]-km values of Paper VI, when 
referred to the component separation and adjusted for 
a ``mean'' $s$$\sim$0.6.  If, however, inner- and 
outer-cone emission occurs along the ``null'' and  
``closed field'' surfaces, then $\rho_{\rm inner[outer]}$
could be some 150[190] km.  While the $h_{\rm inner[outer]}$ 
values appear independent of $P$, only the latter are 
independent of $f$, so that the two emission zones 
may overlap at high frequency---just as is observed 
in the ``boxy'' profiles of many double-cone {\bf M} 
stars above 1 GHz.}

\section{Do the Component Widths Reflect the Conal Geometry?}

We have seen above that, overall, the conal component 
widths exhibit a surprisingly constant spectral behavior.  
However, we also see some possibly significant variations, 
and we here explore whether these can be understood in 
terms of the overall conal emission geometry.

\begin{figure*}
\begin{center}
\begin{tabular}{@{}lr@{}}
{\mbox{\epsfig{file=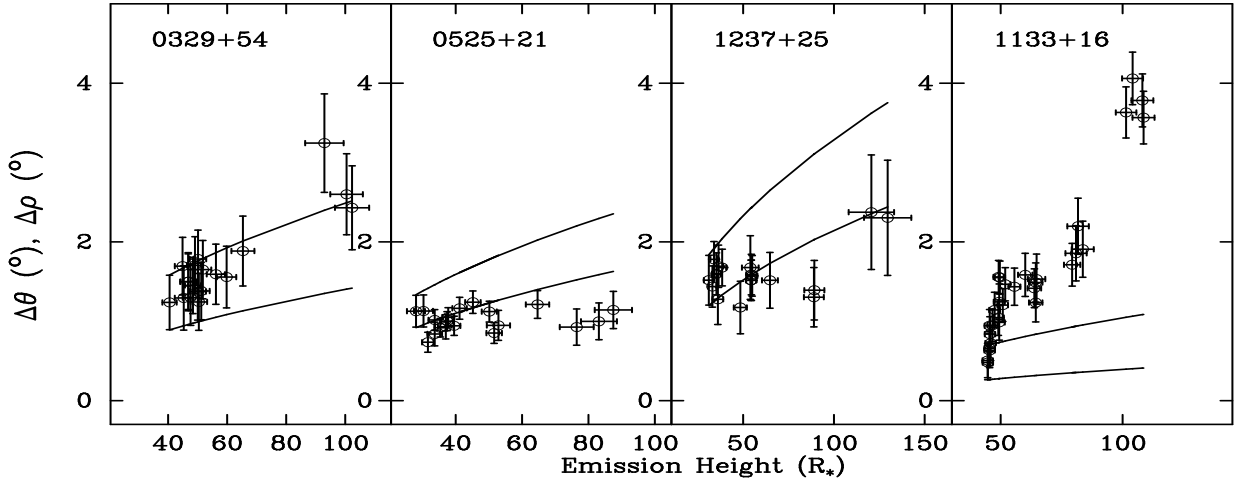,height=16.5cm,width=6.5cm,angle=-90}}}
\end{tabular}
\end{center}
\caption[]{Conal beam width $\Delta\rho$ {\it vs.} emission 
height $h$ for four pulsars with outer-cone component pairs. 
The pairs of curves represent the expected dipolar flaring 
$\Delta\theta$ of a conical emission beam corresponding to 
a polar cap annulus by $\Delta s$ (see text).  $\Delta\rho$ 
and $\Delta\theta$ have errors in both width and $h$ which 
are shown explicitly or included in the bounding curves.} 
\label{fig:flare}
\end{figure*}

At any given frequency, we can use the 3-db width  
together with the component separation to compute 
the {\em radial} half width of the conal emission 
beam $\Delta\rho$ as 
\begin{equation}
\Delta\rho = 2(\rho_{\rm 3db} - \rho_{\rm cs}) ,  
\label{eq9}
\end{equation}
where $\rho_{\rm cs}$ is the beam radius computed 
using the separation and $\rho_{\rm 3db}$ that from 
the 3-db widths.   

Assuming both that $h$ is a monotonic function of 
$f$ and that a given component is associated with 
a specific (uniformly illuminated) bundle of field 
lines (perhaps because a cone of ``emission columns'' 
connects the stellar surface to the emission region 
along them) at a given height, we can ask whether 
the flaring of the (presumed) dipolar magnetic field 
is sufficient to be observed.  In this geometry, the 
angle $\theta$ between the annular emission region 
and the magnetic axis can be expressed in terms of 
the height $h$ as $\theta=1.23^{\circ}s(h/10P)^{1/2}$.  
Then, we can find $\Delta\theta$, the radial 
(half-power) thickness of a thin annulus of ``active'' 
field lines at height $h$ as
\begin{equation}
\Delta\theta=2.45^{\circ}\Delta s(h/10P)^{1/2}
\label{eq10}
\end{equation}

Now we construct a model curve using eq.(\ref{eq10}), 
where $\Delta s$ is estimated at the highest available 
frequency---relative to a nominal $s$ value of 0.6---by 
matching $\Delta\theta$ to $\Delta\rho$.  Then, keeping  
$\Delta s$ constant, $\Delta\theta$ can be calculated 
using eq.(\ref{eq10}) where the height $h$ is obtained 
from the component separation and using eq.(\ref{eq8}) 
with $s$=0.6.  On the other hand, $\Delta\rho$ can 
now be estimated using eq.(\ref{eq9}) and this value 
compared with the corresponding dipolar flaring angle 
$\Delta\theta$ to study the deviations from a dipolar 
geometry. 

In Figure~\ref{fig:flare} we plot both the $\Delta\theta$ 
model and the $\Delta\rho$ widths versus $h$ for most 
of the outer-cone stars.  (B0301+19 was omitted because 
of its strange ``step'', and B2020+28 and B2045--16 owing  
of their limited $h$ ranges.)  The curves indicate the 
dipole-flaring model ($\Delta\theta$) bounded by 
$\Delta s$$\pm$$\sigma_{\Delta s}$, whereas the open 
circles ($\Delta\rho$) show what flaring is seen in the 
observed fits.  Both quantities entail errors in both 
$h$ and $\rho$, which are included in the bounding  
curves ($\sigma_{\Delta s}$) or explicitly indicated.  
For most of the pulsars (including the excluded B2020+28 
and 2045--16) the agreement between these flaring 
estimates is rather good.  For B0329+54 the model works 
particular well, which in turn is compatible with the 
A/R analyses of M\&S and G\&G.  Thus, any component 
narrowing at the highest frequencies can be understood 
as an angular narrowing in the radial (magnetic 
``zenith angle'' $\theta$) thickness of the emitting 
annulus (or cone) of field lines---though in some cases 
this may be compensated by the obliqueness of the 
sightline ``cut'' across the cone as $\beta/\rho$ 
increases with frequency.  Note that we have assumed 
here that $s$ is independent of $f$. This is only true 
if the emission cone is uniformly illuminated at all 
frequencies, and scales in the way given by 
eq.(\ref{eq10}).  Deviations from either of the 
assumptions would lead to a frequency dependent $s$.  
This effect may be refected in Fig.~\ref{fig:flare}, 
where the observed flaring (open circles) is 
consistently either lower or higher than the computed 
flaring (shaded) region.  B1133+16 is the only star 
for which there is a significant discrepancy, possibly 
owing to its remarkably large value of $\beta/\rho$.  
Ignoring the lowest frequency (25 MHz) heights (where 
some scattering or instrumental effects may contribute), 
the width escalation at the lowest heights is much too 
steep to be dipole flaring; whereas, between 50 and 
100 $R_*$, the $s$=0.6 model works well.  

{\it In summary, assessment of whether these outer-cone 
beams flare in a dipolar manner depends critically on 
quality low frequency (large height) observations, which 
often remain problematic.  Though the $s$=0.6 flaring 
model works well (except for B1133+16 as above), the 
B0525+21 analysis suggests a flaring which is {\rm less 
than} dipolar.}  Higher quality observations at the 
lowest possible frequencies are needed to resolve the 
flaring issue.

\section{Discussion and Conclusions}

In the foregoing sections we have attempted an analysis 
of the conal component-pair dimension information for a 
set of 10 pulsars, the seven exemplars previously studied 
by Thorsett (1991) as well as three other prominent stars 
with very different characteristics.  In carrying out our 
analysis, we have placed our focus on the angular scales 
of the (presumed circularly symmetric) conal emission beam 
pattern, not, as have most authors before us, on the 
longitude dimensions of the profiles.  Computation of the 
beam sizes requires geometrical information which can 
perhaps yet only be estimated approximately; however, for 
most of these stars the very different approaches of Lyne 
\& Manchester (1988) and one of us (Paper VI) give quite 
similar results.  

In assessing what may be the most extensive set of well 
measured conal component-pair widths, we find that overall 
these widths vary little with frequency.  We do observe some 
gentle narrowing at high frequencies in one or both of the 
components in some stars, a feature which was also observed 
by Sieber \etal\ (1975). We also find that some components 
appear to broaden at the very lowest frequencies, but here 
either instrumental effects or scattering cannot yet be 
ruled out.  It will be interesting, as coherent dedispersion 
techniques are more regularly applied to very low frequency 
observations over the next few years, to explore this question 
in further detail.  

Because of these various sources of weak variation in the 
component widths, the profile-width measures which include 
them---the half-power (3-db) or 10\%-width measures---we 
find, are more or less problematic indicators of the 
profile-scale variations.   Therefore, our profile-scale 
analyses are based on the component-separation measure.  

Our fitting, using the modified Thorsett relation 
eq.(\ref{eq4}), of the conal beam radii $\rho$ results in 
three strongly contrasting types of behavior.  Four of the 
seven Thorsett stars exhibit continuously varying component 
separation, where, remarkably, the best-fit values of 
$\rho_{\circ}$ are only 2--3-times larger than the angular 
radius of the polar cap $\rho_{\rm pc}$ and the index $a$ 
takes values near --0.45.  Three other stars have component 
separation values which seem to ``saturate'' at very high 
frequencies.  Their best-fit values of $\rho_{\circ}$ are 
3--4 $\rho_{\rm pc}$ and the $a$ values are substantially 
steeper.  Then, in completely contrast to both these groups, 
we find three stars which exhibit almost no component-separation 
variation at all.  

The first two behaviors are apparently exhibited by 
stars with outer-cone component pairs; whereas, the 
three examples of near constancy are all plausible or 
demonstrable examples of inner-cone component pairs.  In 
regard to the first two groups, the A group can be well 
fitted by the ``plasma frequency'' index value of --1/3, 
but the B group cannot---and we have been unable to 
identify any geometrical circumstance that might account 
for the differences.  We have also entertained whether 
higher-order contributions to the magnetic field 
configuration might account for them---but, if so, why 
should we see two types of behavior rather than many?  
Certainly ten stars by no means provides an adequate 
statistical sample, but a preliminary examination of 9 
further pulsars suggests that these too divide between  
the three groups.\footnote{These objects, B0148--06, 
B0402--61, B0523+11, B0751+32, B1737+13, B1821+05, B1857--26, 
B2044+15 and B2319+60, constitute a promising short list 
for further study.}

Furthermore, while the overall success of the modified 
Thorsett relation in fitting the profile-width spectra 
is satisfying, it remains merely an empirical relation.  
How then are we to understand the very necessity of the 
$\rho_{\circ}$ term?  And how could the index $a$ of the 
Thorsett equation have physical implications similar to 
those attributed to simple power-laws of the same index?  
 
A somewhat simpler picture emerged when we fitted the 
separation-determined $h$ values as a direct function 
of $f$ using eq.(\ref{eq7}).  The stars of both groups 
A and B could be fitted well with this relation, though 
the indices $b$ for the latter group were substantially 
steeper.  For the A group it was possible to impose an 
$h_{\circ}$ value of 10 km ($R_*$), without altering 
$\chi^2$ very much.  The real surprise, however, was 
that both groups could be fitted exceedingly well with 
an imposed index value of --2/3.  These latter fits 
(see Table 4) have $h_{\circ}$ values of between 5 
and 15 times $R_*$.  

Noting that emission heights computed from half-power 
widths in relation to the ``last open field lines''---as 
in Paper VI---are not very physical, we suggest that 
more realistic values will come from the use of the 
component separation in relation to an annulus of 
``active'' field lines about half way out on the 
polar cap, $s$$\sim$0.5.  Such considerations lead to 
the conclusion that the 1-GHz outer-cone emission comes 
from a height of around 500--600 km (as opposed to 220 
km in Paper VI), whereas the inner-cone emission stems 
from a height of some 400--500 km ({\it c.f.} 130 km).  
The first range agrees well with the several 
independent emission-height determinations that are 
based on relativistic effects---which we summarized 
for those stars in our groups.  

We have attempted to understand the component-width 
variations in terms of spectral changes in $\beta/\rho$ 
and the height-dependent flaring of a dipolar magnetic 
field.  The analysis seems largely consistent with 
such flaring (but depends critically on the difficult 
low frequency observations), and one pulsar suggests 
no flaring at all.  Clearly, our study lends no simple 
support to the premise that the leading members of a 
component pair will be broadened by aberration (Ahmadi 
\& Gangadhara 2002).

We are uncertain about how the strength and structure  
of the pulsar magnetic field might affect the radio 
emission heights. It may be that the magnetic field 
configuration in the emission region will usually be 
nearly dipolar, and that closer to the surface there 
will always be non-dipolar effects [as the multipole 
field will not decay over the pulsar's lifetime 
(Mitra \etal\ 1999)]. This will influence the gap 
heights in all vacuum-gap models---implicitly in the 
R\&S theory and explicitly in later work as by, for 
instance, Muslimov \& Tsygan (1992) or Gil \& Mitra 
(2001).  (It will also complicate interpretation of 
the $s$ parameter.)  Further, non-dipolar fields will 
increase the binding energy of the ions on the stellar 
surface (Abrahams \& Shapiro 1991, Gil \& Mitra 2001), 
which is by far the most problematic issue in forming 
gaps in pulsars. 

In several theories we might expect the gap height 
to have an inverse relationship with the $B$-field 
strength, and thus, the emission region which is 
argued to be proportional to the gap height, may 
start at a lower altitude in pulsars with stronger 
fields; however, here the field in question must be 
the total field in the gap region, including both 
the dipolar and non-dipolar contributions.  We note 
that two of the A-group stars have (spindown-determined, 
meaning dipolar?) field strengths of $\sim$10$^{13}$ 
Gauss, while none of the B- or C-group star fields are 
this high (see Table~\ref{tbl-1}).  That many 
interpulsars seem to emit from both poles almost 
equally is also worth remembering, in that it implies 
that gaps can be formed above polar caps of either 
sense (with accelerated charges of either sign?)---and 
perhaps the two gap senses give rise to the two RFM 
types (A and B) observed above.     

A major impetus for this work was that of trying to 
forge a better overall understanding of ``RFM mapping'' 
to facilitate its physical interpretation.  The summary 
of Xilouris \etal\ is very useful, but we have been 
unable to locate any full theoretical discussion of RFM 
[excepting that of Barnard \& Arons (1986), which predicts 
no RFM  at all].  Returning to R\&S, we find  that those 
stars which exhibit RFM---that is, our groups A and B---can 
be fitted with an index of --2/3, which they associated 
with curvature radiation.  Of course, our fitting does 
not {\it determine} this index, as it is highly correlated 
with the other two parameters; however, it is significant, 
we believe, that an index of --2/3 can be imposed on both 
the group A and B stars without any serious degradation 
of $\chi^2$.  

If the index $b$ in the modified Thorsett relation can be 
interpreted straightforwardly in R\&S terms, then this 
dependence represents an underlying physical situation 
in which particles of a particular $\gamma$ radiate at 
different heights because the field-line curvature 
$\rho_c$ and the plasma frequency $\omega_p$ in turn vary 
systematically with height.  Thus, if the magnetic field 
is dipolar in the emission region, perhaps R\&S's eq.(64) 
can meaningfully be evaluated to obtain the controlling 
quantity $\gamma_{\rm max}$. Other theories, however, may 
view this circumstance very differently, but must find a 
natural explanation for this specific RFM phenomenon.  

A nearly opposite circumstance is observed in stars with
no RFM.  How can broad-band emission at a single, fixed
height be understood?---a situation which R\&S have not
at all addressed.  Apparently, this inner-cone ``no-RFM''
situation is as general as the outer-cone RFM, and indeed
both situations commonly appear in those stars with both
inner- and outer-cone emission.  This dichotomy is 
suggestive of the wave-mode propagation model first 
advanced in comprehensive detail by Barnard \& Arons (1986). 
Though not free of problems, this model addresses the 
effects of birefringence in the pulsar magnetosphere, 
such that one propagating wave-mode escapes the magnetosphere 
at different heights mimicking the usual (outer cone) RFM 
behavior, whereas the other mode can probably escape from 
a fixed height with results similar to the ``no-RFM'' (inner 
cone) phenomenon.  This said, it is not clear that inner and 
outer cones exhibit the systematic polarization differences 
which might be expected of distinct propagating wave modes---and,
birefringence could also be involved in the angular shifts 
seen between the two polarization modes (Rankin \& Ramachandran 
2002).  So, while we cannot be sure that propagation effects 
are responsible for these dramatic differences, we are sympathetic 
to the possibility, and it remains possible that a more flexible 
propagation model such as Petrova's (2001 and the references 
cited therein) could account for the systematics of both  
beaming and polarization.

So, in conclusion, our results seem nominally compatible 
with the picture that conal beams are produced by a 
system of subbeams circulating around the magnetic axis 
some halfway out on the polar cap, which in turn are 
produced, perhaps, by excitation in ``vaccum gaps'' and 
radiate by a mechanism wherein the plasma frequency 
determines the emission height.  Nothing, however, in 
our analysis, has demonstrated these specific physical 
circumstances.

\acknowledgments

We thank Steve Thorsett for generously sharing his notes and 
the profile measurements used in the preparation of his earlier 
paper.  We also thank Michael Kramer both for his insightful 
comments on our manuscript as well as for the use of his 
Gaussian-fitting programs, and Yashwant Gupta for providing 
several needed profiles using the GMRT.  
Several profiles were obtained from the extremely useful 
pulsar database of the European Pulsar Network maintained 
by Max-Planck Institut f\"ur Radioastronimie.
We also learned much from 
discussions and comments on early versions of the manuscript 
from Jim Cordes, Jan Gil, Aris Karastergiou, George Melikidze, 
R. Ramachandran, Mal Ruderman, C. S. Shukre, Axel Jessner and Steve Thorsett.  
Portions of this work were carried out under US National Science 
Foundation (NSF) Grants INT 93-21974 and AST 99-87654 as well as 
under a visitor grant from the Nederlandse Organisatie voor 
Wetenschappelijk Onderzoek.  Arecibo Observatory is operated by 
Cornell University under contract to the US NSF.



\begin{deluxetable}{rcccccc}
\tablewidth{0pt}
\tablecaption{Fit Parameters for the Leading and Trailing Components Widths.\label{tbl-2}}
\tablehead{
Pulsar B-/ &Freq-range&   $K$    &    $\zeta$     & $\chi^{2}$ &  M & References  \\
Component  &   (GHz)  &($\deg/{\rm GHz}^{\zeta}$)& &          &   &    }
\startdata
 0301+19 I & 0.1--4.9 &    3.7$\pm$1.6   &  0.20$\pm$0.02 &  1.20      &19&  1,2,3   \\
        II & 0.1--4.9 &    3.9$\pm$1.0   &  0.04$\pm$0.02 &  1.00      &19&  4,6,9   \\
         I & 0.1--0.5 &    2.7$\pm$0.1   & -0.05$\pm$0.09 &  0.16      &9 &  10,11   \\
        II & 0.1--0.5 &    4.6$\pm$0.1   &  0.14$\pm$0.09 &  1.60      &9 &  14      \\
         I & 0.6--4.9 &    4.1$\pm$0.1   &  0.09$\pm$0.05 &  1.01      &10&          \\
        II & 0.6--4.9 &    4.1$\pm$0.2   &  0.04$\pm$0.06 &  0.60      &10&          \\
           &          &          &                &            &  &          \\
 0329+54 I &0.1--10.5 &    2.9$\pm$0.01   &  0.01$\pm$0.03 & 0.4        &19&  2,4,5   \\
        II & 0.1--10.5&    3.2$\pm$0.02   & -0.21$\pm$0.02 & 1.0        &19&  7,8,9  \\
        II & 0.2--10.5&    3.0$\pm$0.02   & -0.12$\pm$0.05 & 0.5        &16&  10.11      \\
           &          &          &                &            &  &          \\
 0525+21 I &0.1--4.9  &    2.7$\pm$0.01   & -0.06$\pm$0.03 & 0.76       &19&  2,3,4,9   \\
        II &0.1--4.9  &    2.8$\pm$0.01   & -0.04$\pm$0.03 & 1.69       &19&  10,11,14 \\
           &          &          &                &            &  &          \\
 1133+16 I &0.02--10.5&    2.1$\pm$0.03   & -0.16$\pm$0.01 & 2.7        &31&  2,3,9   \\
        II &0.02--10.5&    3.2$\pm$0.01   & -0.13$\pm$0.01 & 5.5        &31&  4,6,7   \\
         I &0.04--10.5&     2.2$\pm$0.01  & -0.06$\pm$0.02 & 1.14       &26&  10,11   \\
        II &0.04--10.5&     3.3$\pm$0.01  & -0.09$\pm$0.02 & 5.00       &26&          \\
           &          &          &                &            &  &          \\
 1237+25 I &0.04--4.9 &    1.7$\pm$0.02   & -0.005$\pm$0.03&  1.1       &18&  1,2,3,4  \\
        II &0.04--4.9 &    2.2$\pm$0.01   &  0.006$\pm$0.02&  2.1       &18&  9,10,11,14   \\
           &          &          &                &            &  &          \\
 2020+28 I &0.1--10.5 &     3.4$\pm$0.01  & -0.04$\pm$0.03 & 4.0        &15&  2,3,9  \\
        II &0.1--10.5 &     2.7$\pm$0.01  & -0.13$\pm$0.03 & 1.8        &15&  4,10,11,12 \\
           &          &          &                &            &  &          \\
 2045--16 I &0.1--4.9 &    2.9$\pm$0.01   & -0.14$\pm$0.04 & 1.5        &12&  2,6,4  \\
        II &0.1--4.9  &    2.6$\pm$0.01   & -0.19$\pm$0.04 &  1.5       &12&  10,11,13  \\ 
           &          &          &                &            &  &          \\
 0834+06 I &0.04--4.9 &    2.1$\pm$0.02   &  0.04$\pm$0.02 & 1.5        &17&  3   \\
        II &0.04--4.9 &    2.9$\pm$0.1    & -0.01$\pm$0.01 & 3.7        &17&      \\
           &          &          &                &            &  &          \\
 1604--00 I&0.04--4.9 &    2.1$\pm$0.3   & -0.09$\pm$0.05 & 3.0        & 6&  3   \\
        II &0.04--4.9 &    1.4$\pm$0.2   & -0.15$\pm$0.05 & 1.5        & 6&      \\
           &          &          &                &            &  &          \\
 1919+21 I &0.04--4.9 &    3.1$\pm$0.1    &  0.10$\pm$0.04 & 1.1        &15&  3   \\
         I &0.4--4.9  &    3.3$\pm$0.02  & -0.10$\pm$0.09 & 3.0        &11&      \\
        II &0.04--4.9 &    3.2$\pm$0.01   & -0.01$\pm$0.03 & 1.0        &15&      \\ 
        II &0.4--4.9  &    3.2$\pm$0.01   & -0.11$\pm$0.05 & 0.4        &11&      \\ \tableline\enddata
\tablerefs{
(1) Arzoumanian \etal\ 1994;
(2) Gould \& Lyne 1998;
(3) Hankins, Eilek \& Rankin 2002;
(4) Hoensbroech \& Xilouris 1997a:
(5) Hoensbroech \& Xilouris 1997b;
(6) Kijak \etal\ 1997;
(7) Kramer \etal\ 1997;
(8) Kuz'min \& Izvekova 1996;
(9) Kuz'min \& Losovskii 1999;
(10) Seiradakis \etal\ 1995;
(11) Thorsett 1991;
(12) Gupta 2001 ({\em private communication});
(13) McCulloch \etal\ 1982;
(14) Mitra 2002 ({\em Unpublished test observations at 8.35 GHz using Effelsberg
     Radio Telescope, private communication.})}

\end{deluxetable}

\begin{deluxetable}{ccccccc}
\tablewidth{0pt}
\tablecaption{Fit Parameters for the Conal Beam Radii.\label{tbl-3}}
\tablehead{
\colhead{Pulsar (B--)}& \colhead{$\rho_{\circ}(\deg)$} &
\colhead{$\rm{f}_\circ$ (GHz)}          & \colhead{$a$}&
\colhead{$C_{ij}$ (\%)}          & \colhead{$\chi^{2}$}     &
\colhead{$M$}}
\startdata
\sidehead{Component Separation --- Group A}
0301+19     & $2.29\pm0.2$ & $2.1\pm0.9$           &-0.47$\pm$0.06  &+92,-96,-99& 0.5 & 29 \\
0525+21     & $1.70\pm0.2$ & $0.4\pm0.01$          &-0.40$\pm$0.08  &+93,-97,-99& 0.5 & 19 \\
1237+25     & $2.98\pm0.2$ & $1.1\pm0.2$           &-0.45$\pm$0.05  &+90,-95,-98& 0.5 & 42 \\
2045--16    & $2.72\pm0.2$ & $0.4\pm0.05$          &-0.42$\pm$0.01  &+93,-97,-99& 1.0 & 32 \\[4pt]
0301+19     & {\bf 1.040}  & $48\pm5$              &-0.26$\pm$0.01  &  -98.7    & 0.9 & 29 \\
0525+21     & {\bf 0.633}  & $30\pm10$             &-0.18$\pm$0.01  &  -98.8    & 1.1 & 19 \\
1237+25     & {\bf 1.042}  & $390\pm100$           &-0.19$\pm$0.01  &  -98.5    & 1.1 & 42 \\
2045--16    & {\bf 0.875}  & $552\pm200$           &-0.15$\pm$0.01  &  -98.7    & 1.4 & 32 \\[4pt]
0301+19     & {\bf 1.040}  & $19.5\pm0.3$          & {\bf -1/3}     &   ---     & 2.2 & 29 \\
0525+21     & {\bf 0.633}  & $5   \pm0.5$          & {\bf -1/3}     &   ---     & 20  & 19 \\
1237+25     & {\bf 1.042}  & $20  \pm0.3$          & {\bf -1/3}     &   ---     & 13  & 42 \\
2045--16    & {\bf 0.875}  & $13  \pm0.5$          & {\bf -1/3}     &   ---     & 28  & 32 \\[4pt]
0301+19     & $1.59\pm0.3$ & $ 10.2\pm1.2$         & {\bf -1/3}     &  -94.3    & 0.7 & 29 \\
0525+21     & $1.5\pm0.1$  & $ 0.7\pm0.3$          & {\bf -1/3}     &  -94.3    & 0.6 & 19 \\
1237+25     & $2.4\pm0.3$  & $ 3.9\pm0.5$          & {\bf -1/3}     &  -93.7    & 0.6 & 42 \\
2045--16    & $2.34\pm0.3$ & $1.2 \pm0.4$          & {\bf -1/3}     &  -93.9    & 1.1 & 32 \\
\sidehead{Component Separation --- Group B}
0329+54     & $6.0\pm0.1$  & $0.259\pm0.045$       &-1.07$\pm$0.17  &+83,-88,-99& 0.9 & 19 \\
1133+16     & $4.4\pm0.02$ & $0.121\pm0.033$       &-0.55$\pm$0.03  &+74,-83,-97& 1.1 & 62 \\
2020+28     & $8.7\pm0.1$  & $0.165\pm0.080$       &-0.89$\pm$0.11  &+76,-75,-99& 1.0 & 35 \\[4pt]
0329+54     & $4.6\pm0.3$  & $3.4  \pm1.0  $       & {\bf -1/3}     &  -90.1    & 1.6 & 19 \\
1133+16     & $3.9\pm0.2$  & $0.4  \pm0.3  $       & {\bf -1/3}     &  -86.0    & 5.6 & 62 \\
2020+28     & $8.3\pm0.3$  & $0.381\pm1.4  $       & {\bf -1/3}     &  -91.0    & 3.0 & 35 \\[4pt]
\sidehead{Half-power Widths --- Group A}
0301+19     & {\bf 1.040}  & $1.8\pm0.5\times10^3$ &-0.185$\pm$0.03 &  -98.7    & 1.3 & 29 \\
0525+21     & {\bf 0.633}  & $262\pm100$           &-0.15$\pm$0.01  &  -98.7    & 1.0 & 19 \\
1237+25     & {\bf 1.042}  & $9.6\pm5.0\times10^3$ &-0.15$\pm$0.02  &  -98.3    & 1.6 & 18 \\
2045--16    & {\bf 0.875}  & $5.0\pm3.0\times10^3$ &-0.14$\pm$0.02  &  -99.3    & 1.2 & 12 \\[4pt]
\sidehead{Half-power Widths --- Group B}
0329+54     & $6.4\pm0.4$  & $0.359\pm0.085$       &-1.00$\pm$0.22  &+83,-88,-99& 0.7 & 19 \\
1133+16     & $4.6\pm0.2$  & $0.310\pm0.100$       &-0.51$\pm$0.30  &+75,-84,-98& 4.0 & 30 \\
2020+28     & $9.4\pm0.3$  & $0.205\pm0.160$       &-0.88$\pm$0.15  &+85,-89,-99& 0.6 & 16 \\[4pt]
\sidehead{10\% Widths --- Group A}
0301+19     & {\bf 1.040}  & $2.6\pm0.4\times10^5$ &-0.13$\pm$0.02  &  -99.7    & 0.6 & 24 \\
0525+21     & {\bf 0.633}  & $2.9\pm1.2\times10^3$ &-0.13$\pm$0.02  &  -99.1    & 1.1 & 17 \\
1237+25     & {\bf 1.042}  & $1.6\pm1.5\times10^3$ &-0.13$\pm$0.02  &  -99.3    & 0.5 & 20 \\
2045--16    & {\bf 0.875}  & $1.0\pm1.0\times10^5$ &-0.12$\pm$0.03  &  -99.5    & 0.5 &  8 \\[4pt]
\sidehead{10\% Widths --- Group B}
0329+54     & $7.4\pm0.4$  & $0.401\pm0.200$       &-1.12$\pm$0.30  &+75,-78,-99& 0.5 & 18 \\
1133+16     & $5.3\pm0.2$  & $0.350\pm0.080$       &-0.55$\pm$0.02  &+93,-98,-98& 3.5 & 36 \\
2020+28     & $9.9\pm0.7$  & $0.544\pm0.350$       &-0.83$\pm$0.10  &+90,-94,-99& 0.7 & 15 \\[4pt]
\sidehead{Component Separation --- Group C}
0834+06     & $3.6\pm0.3$  &    ---                &    ---         &   ---     & 0.2 & 17 \\
1604-00     & $6.1\pm0.4$  &    ---                &    ---         &   ---     & 0.3 &  6 \\
1919+21     & $3.3\pm0.2$  &    ---                &    ---         &   ---     & 0.2 & 15 \\[4pt]
\sidehead{Half-power Widths --- Group C}
0834+06     & $3.9\pm0.3$  &    ---                &    ---         &   ---     & 0.1 & 17 \\
1604-00     & $6.8\pm0.4$  &    ---                &    ---         &   ---     & 0.1 &  6 \\
1919+21     & $3.8\pm0.3$  &    ---                &    ---         &   ---     & 0.8 & 15 \\[4pt]
\sidehead{10\% Widths --- Group C}
0834+06     & $4.2\pm0.3$  &    ---                &    ---         &   ---     & 0.3 & 16 \\
1604-00     & $8.1\pm0.3$  &    ---                &    ---         &   ---     & 1.6 &  6 \\
1919+21     & $4.2\pm0.5$  &    ---                &    ---         &   ---     & 0.1 & 15 \\[4pt]
\enddata
\end{deluxetable}

\begin{deluxetable}{ccccccc}
\tablewidth{0pt}
\tablecaption{Fit Parameters for the Emission Heights.\label{tbl-4}}
\tablehead{
\colhead{Pulsar (B--)}& \colhead{$\rm{h}_{\circ}(Km)$} &
\colhead{$\rm{f_h}$  (GHz)}          & \colhead{$b$}&
\colhead{$C_{ij}$ (\%)}          & \colhead{$\chi^{2}$}     &
\colhead{$M$}}
\startdata
\sidehead{Component Separation --- Group A}
0301+19     & $57.7\pm7.0$ & $365\pm50 $           &-0.69$\pm$0.07  &+89,-93,-99& 0.5 & 29 \\
0525+21     & $81.0\pm12.8$& $2.1\pm1.0\times 10^3$&-0.54$\pm$0.08  &+91,-95,-99& 0.5 & 19 \\
1237+25     & $91.9\pm8.1$ & $779\pm100$           &-0.61$\pm$0.05  &+87,-92,-99& 0.5 & 42 \\
2045--16    & $104\pm11.0$ & $1.8\pm0.9\times 10^3$&-0.54$\pm$0.08  &+91,-95,-99& 1.0 & 32 \\[4pt]
0301+19     & {\bf 10}     & $8.7\pm5.0\times 10^3$&-0.42$\pm$0.02  &   98.9    & 1.1 & 29 \\
0525+21     & {\bf 10}     & $2.2\pm0.7\times 10^7$&-0.29$\pm$0.03  &   98.9    & 1.1 & 19 \\
1237+25     & {\bf 10}     & $1.4\pm0.6\times 10^7$&-0.30$\pm$0.02  &   98.6    & 1.2 & 42 \\
2045--16    & {\bf 10}     & $1.7\pm0.8\times 10^9$&-0.24$\pm$0.02  &   98.8    & 1.4 & 32 \\[4pt]
0301+19     & $54.7\pm5.0$ & $5.0\pm2.0\times 10^3$& {\bf -2/3}     &  -81.0    & 0.5 & 29 \\
0525+21     & $95.2\pm7.0$ & $360\pm50$            & {\bf -2/3}     &  -79.8    & 0.6 & 19 \\
1237+25     & $98.0\pm10.0$& $380\pm200$           & {\bf -2/3}     &  -78.8    & 0.5 & 42 \\
2045--16    & $117.1\pm10.0$&$320\pm50$            & {\bf -2/3}     &  -78.7    & 1.0 & 32 \\
\sidehead{Component Separation --- Group B}
0329+54     & $158.7\pm10.0$&$43.8\pm10.0  $       &-0.87$\pm$0.17  &+80,-84,-99& 0.5 & 19 \\
1133+16     & $158.9\pm10.0$&$29.9\pm5.03  $       &-0.76$\pm$0.03  &+70,-79,-98& 0.6 & 62 \\
2020+28     & $178.6 \pm10.0$&$4.3 \pm1.0   $       &-1.10$\pm$0.11  &+73,-77,-99& 0.8 & 35 \\[4pt]
0329+54     & $147.3\pm10.0$&$302\pm50     $       & {\bf -2/3}     &  -69.0    & 0.6 & 19 \\
1133+16     & $153.6\pm10.0$&$85.5\pm10.0  $       & {\bf -2/3}     &  -64.0    & 0.9 & 62 \\
2020+28     & $173.5\pm5.0$  &$45.6\pm5.0   $       & {\bf -2/3}    &  -73.0    & 1.2 & 35 \\
\enddata

\end{deluxetable}
\end{document}